\documentclass[10pt,pre,a4paper,twocolumn,superscriptaddress,showpacs]{revtex4-2}
\usepackage[utf8]{inputenc}
\usepackage[colorlinks=true]{hyperref}
\usepackage{amsmath,amssymb}
\usepackage{amsfonts}
\usepackage[dvips]{graphicx}
\usepackage[dvips]{color}
\usepackage[dvipsnames]{xcolor}
\usepackage{epstopdf}

\begin{document}

\title{Structure of the Lennard-Jones liquid estimated from a single simulation}
\author{Shibu Saw}\email{shibus@ruc.dk}\affiliation{``Glass and Time,” IMFUFA, Department of Science and Environment, Roskilde University, P.O. Box 260, DK-4000 Roskilde, Denmark}
\author{Jeppe C. Dyre}\email{dyre@ruc.dk}\affiliation{``Glass and Time,” IMFUFA, Department of Science and Environment, Roskilde University, P.O. Box 260, DK-4000 Roskilde, Denmark} 

\date{\today}

\begin{abstract}{
		Combining the recent Piskulich-Thompson approach [Z. A. Piskulich and W. H. Thompson, {J. Chem. Phys.} {\bf 152}, 011102 (2020)]  
		with isomorph theory, from a single simulation, the structure of a single-component Lennard-Jones (LJ) 
		system is obtained  at an arbitrary state point in almost the whole liquid region of the 
		temperature-density phase diagram. The LJ system exhibits two temperature range 
		where the van't Hoff's assumption that energetic and entropic forces are temperature independent is valid. 
		A method to evaluate the structure at an arbitrary state point along an isochore from the knowledge 
		of structures at two temperatures on the isochore is also discussed. 
		{We argue that, in general, the 
		structure of any hidden scale-invariant system obeying the van't Hoff's assumption in the whole range of 
		temperatures can be determined in the whole liquid region of the phase diagram from only a single 
		simulation.}
   }
\end{abstract}

\pacs{xx}

\maketitle

\section{Introduction}
The structure of an equilibrium liquid is characterized by the radial distribution function $g(r)$. This quantity can 
be obtained by light scattering experiments, simulation or liquid state theory\cite{hansen-mcdonald,saw2015,saw2015jcp}.
This quantity provides not only an idea of the structure, but also facilitates in predicting various thermodynamics 
quantities as $g(r)$ is related to the latter through interparticle interactions\cite{mcquarrie}. The static structure factor, 
which is the Fourier transform of $g(r)$, is an input of the mode-coupling theory (MCT), which yields dynamical 
quantities such as the mean-squared displacement (MSD) or the intermediate scattering function\cite{kob,gotze}.

Experiments are tricky to perform for supercooled liquids, which have a strong tendency to crystallize.
Simulations are equally difficult and need to be performed for a very long time due to associated long relaxation times\cite{berthier2011}.

Theoretical study of the temperature dependence of the structure will facilitate the prediction of structure
from limited experimental or simulation data; however, such studies are limited\cite{Shen2018}.
Piskulich and Thomson\cite{Thompson} have shown that the radial distribution function $g(r)$ of TIP4P/2005 
water\cite{tip4p-2005} at several temperatures can be obtained from a single simulation. This 
theory is based on the van't Hoff's assumption\cite{van-Hoff} that the energetic and entropic forces are temperature 
independent.

In this paper, we test the van't Hoff's assumption for the single-component Lennard-Jones (LJ) system. 
This assumption is not valid for the whole range of temperatures, but it is valid for two ranges 
of temperatures separately.
The Piskulich-Thompson theory, then, is employed to the LJ system to predict the structure at other temperatures 
along the same isochore in each temperature range separately. We also prescribe a method to predict the 
structure from knowledge of $g(r)$ at two different temperatures along the same isochore, without performing 
any simulation or experiment. 
{A class of systems exhibits a strong correlation between virial and potential energy equilibrium fluctuations, such as 
the inverse power law, LJ, etc., known as {\it Roskilde or R-simple} system. 
Along an isomorph\cite{Dyre2012Review,Dyre2014Review,Dyre2016Review,Dyre2018Review}, an {\it R-simple} system's 
structure and dynamics are invariant in the reduced unit\cite{isomorph2009,isomorph2014}.} 
We combine here the Piskulich-Thompson\cite{Thompson} approach with isomorph theory to predict the structure of the LJ system at 
an arbitrary state point in the liquid region of the temperature-density phase diagram. 

We describe the Piskulich-Thompson theory in Sec.~\ref{theory}. Sec.~\ref{simulation} describes the 
simulation method used. The results are given in Sec.~\ref{result}. The Sec.~\ref{no-sim} 
explains how $g(r)$ along an isochore can be obtained without any simulation if the radial distribution functions 
at two temperatures along the same isochore are known. The extension of the Piskulich-Thompson theory for R-simple liquids is described in Sec.~\ref{pt-iso-theory}. A summary and discussions are given in Sec.~\ref{summary}. 

\section{Piskulich-Thompson theory}\label{theory}
The radial distribution function is defined by\cite{allen-tildesley}
\begin{eqnarray}
	g(r) = \dfrac{V}{N^2} \left < \sum_i \sum_{j \ne i} \delta(r-r_{ij}) \right>, \label{gr}
\end{eqnarray}
where $r_{ij}$ is the distance between particles $i$ and $j$, $V$ and $N$ are the volume and the 
number of particles, respectively. The $<\cdots >$ represents ensemble average.
{
Since $\underset{i}\sum \underset{j \ne i}\sum \delta(r-r_{ij})$ does not depend on the momenta explicitly, Eq.\eqref{gr}
can be re-written as
\begin{eqnarray}
	g(r) = \dfrac{V}{N^2} \dfrac{1}{Z} \int {d\bf{q}}  e^{-\beta {\cal U}} \sum_i \sum_{j \ne i} \delta(r-r_{ij}) , \label{gr2}
\end{eqnarray}
where $d\bf{q}$ are the system coordinates and $\beta=(k_BT)^{-1}$, $k_B$, and $T$ being the Boltzmann constant and temperature, respectively.
$Z$ and ${\cal U}$ are the configurational canonical partition function  and potential energy of the system.
The temperature dependence of the radial distribution function $g(r)$ is given by\cite{Thompson}
\begin{eqnarray}
	\dfrac{\partial g(r)} {\partial \beta} = -\dfrac{V}{N^2} \left < \Delta {\cal U} \sum_i \sum_{j \ne i} \delta(r-r_{ij}) \right> , \label{gH}
\end{eqnarray}
where $\Delta {\cal U} = {\cal U} - <{\cal U}>$ is the fluctuation of potential energy from its mean value $<{\cal U}>$.
It should be noted that ${\cal U}$ is the total potential energy of a configuration, not of an individual particle.
}

The Helmholtz free energy profile $ A(r)$ can be written in terms of the radial 
distribution function $g(r)$ as\cite{Thompson}
\begin{eqnarray}
	 A(r) = -k_B  T \ln g(r) - k_B T \ln \nu(r), \label{A}
\end{eqnarray}
where $\nu(r)=r^2$ is a geometric factor.
Without the geometric factor, the free energy is simply the potential of mean force $F_{PM}(r)$. The 
derivative of the Helmholtz free energy with respect to $\beta$ is
\begin{eqnarray}
	\dfrac {\partial A(r)} {\partial \beta} &=& k_B T \left[ \dfrac {g_H(r)}{g(r)}  + k_B T \ln g(r) + k_B T \ln \nu(r) \right] \label{Aprime-mine} \\
						&=& k_B T \left[ \dfrac {g_H(r)}{g(r)}  - A(r)  \right], \label{Aprime}
\end{eqnarray}
where $g_H(r) \equiv -\dfrac{\partial g(r)} {\partial \beta}$. The Helmholtz free energy $A(r)$ can be written in terms of internal energy and entropy as
\begin{eqnarray}
	 A(r) = U(r) -  T S(r). \label{A2}
\end{eqnarray}
With the assumption that both $U(r)$ and $ S(r)$ do not depend on the 
temperature (van't Hoffian assumption), a comparison of equations \eqref{Aprime} and \eqref{A2} 
yields expression for the internal energy and the entropy as,
\begin{eqnarray}
	U(r) = \dfrac{g_H(r)}{g(r)}, \label{U}
\end{eqnarray}
and
\begin{eqnarray}
	S(r) = \dfrac{1}{k_BT^2} \dfrac{\partial A(r)}{\partial \beta}. \label{S}
\end{eqnarray}
While $ U(r)$ can be readily evaluated from equations \eqref{gr} and \eqref{gH}, 
the entropy $S(r)$ can be determined from equations \eqref{Aprime-mine}, \eqref{gH}, and \eqref{gr}. 
Thus one can calculate the value 
of $ U(r)$ and $S(r)$ from simulation data at a given temperature $T_0$. 
Now from equation \eqref{A}, the radial distribution function at an arbitrary 
temperature $T$, but same density, can be written as
\begin{eqnarray}
	g(r;\beta) = \dfrac{1}{\nu(r)} e^{-\beta U(r)} e^{S(r)/k_B}, \label{grT}
\end{eqnarray}
where $ U(r)$ and $S(r)$ are evaluated at the temperature $T_0$.
The above equation gives rise to the van't Hoff plot\cite{Atkins} 
if $ U(r)$ and $S(r)$ are 
assumed to be temperature independent.

Substituting the values of $U(r)$ from equation \eqref{U} and $\dfrac{\partial A(r)}{\partial \beta}$ from 
equation \eqref{Aprime-mine}, equation \eqref{grT} becomes\cite{Thompson}
\begin{eqnarray}
	g(r;\beta) = g(r;\beta_0) e^{ U(r)(\beta_0-\beta)}. \label{grT-mine}
\end{eqnarray}
This expression of $g(r;\beta)$ depends only on $ U(r) \equiv \dfrac{g_H(r)} {g(r)}$. It must 
be emphasized again that throughout the derivation, the van't Hoffian assumption is assumed.
\section{Simulation Details} \label{simulation}
We have performed a canonical ensemble molecular dynamics simulations (NVT) of the LJ system employing a Nose-Hover 
thermostat with $N=2000$ particles at various densities and 
temperatures. Employing a shifted-forces cutoff\cite{Toxvaerd} the LJ interaction potential between particle $i$ and $j$ is given as
\begin{eqnarray}
	\dfrac{\phi(r_{ij})}{4\epsilon} = \left\{
	\begin{array}{ll}
		(\frac{\sigma}{r_{ij}})^{12} - (\frac{\sigma}{r_{ij}})^6 + C_1 r + C_2, &r_{ij} < 2.5\sigma, \\
			 0, & r_{ij} \ge 2.5\sigma,
	\end{array}
	\right .
	\label{lj}
\end{eqnarray}
where $C_1$ and $C_2$ ensure that the $\phi(r)$ and its first derivative are continuous at the 
cut-off $r=2.5\sigma$. The simulations were performed using RUMD (Roskilde University 
Molecular Dynamics) software\cite{rumd} which is a GPU (graphical processing unit) code. All quantities 
reported in this paper are in LJ units: length, time and temperature are expressed in units of 
$\sigma$, $\sqrt{m\sigma^2/\epsilon}$ and $\epsilon/k_B$, respectively.
{
	We have used the state point dependent molecular dynamics (MD) time step given by $\Delta t = {0.001}\sqrt{m/(k_BT\rho^{2/3})} $.  
	Most of the data result from $5.10^{7}$ steps equilibration followed by $2.10^{8}$ steps production run.  
}

\section{Results}\label{result}
\subsection{The validity of van't Hoffian assumption}
\begin{figure}[t]
	\includegraphics[width=0.40\textwidth,angle=0]{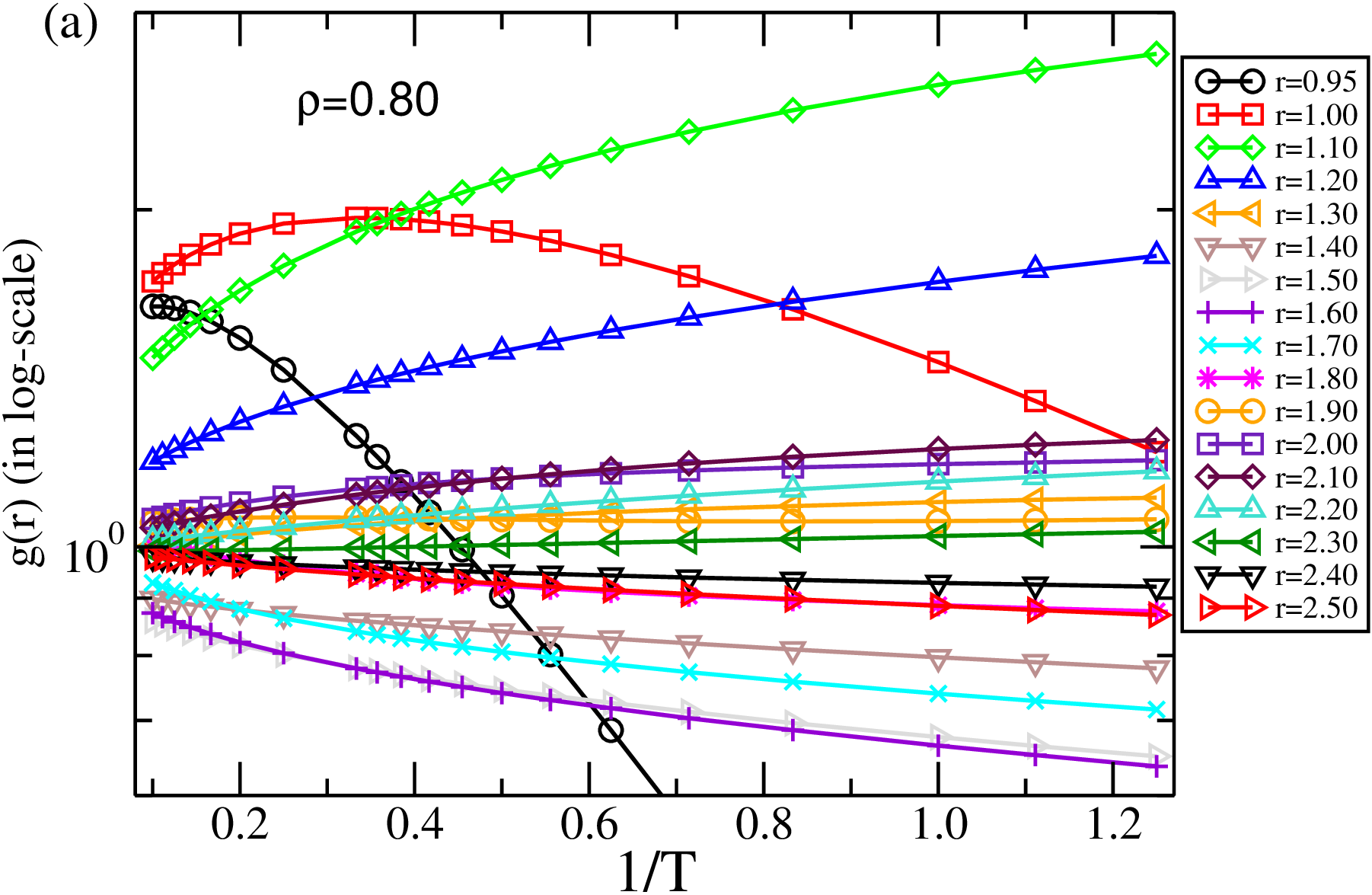}\vspace{10mm}
	\includegraphics[width=0.40\textwidth,angle=0]{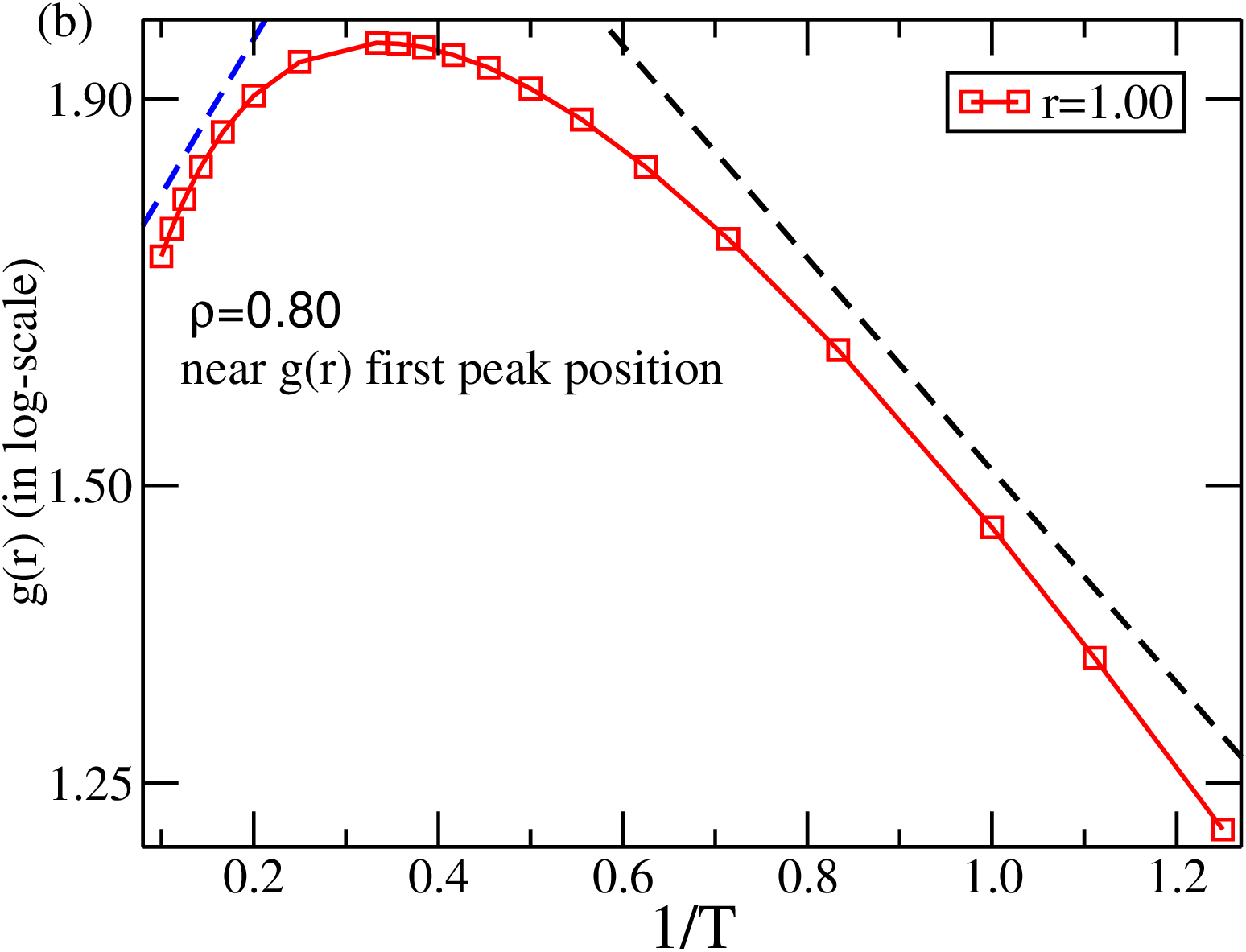}\vspace{0mm}
	\caption{$T^{-1}$-dependence of $g(r)$ of a LJ system at density $\rho=0.80$ in log-linear scale for (a) different values of r and (b) $r = 1.0$. The dashed lines are guides to the eye.}
\label{gr-invT}
\end{figure}

If the van't Hoffian assumption is correct, then from equation \eqref{grT} or \eqref{grT-mine} 
$\ln g(r;\beta)$ vs. $\beta$ should be a straight line. We plot the $g(r)$ against 
$1/T$ in log-linear scale in figure \ref{gr-invT}(a) for the LJ system at density $\rho=0.80$ for a range 
of $r$ values. It  shows that the $\ln g(r)$ vs $1/T$ are not straight lines throughout the 
considered temperature range. This means that the van't Hoffian assumption that $ U(r)$ and 
$S(r)$ are temperature independent is not correct. 
The van't Hoffian assumption has been seen not to be valid in other liquids or liquid mixtures 
with covalent bonds\cite{Dorsey,Galaon}, as well. Interestingly, the van't Hoffian assumption is 
not valid for the LJ system which is a very simple liquid without any covalent bonds. 
However, two temperature ranges can be assigned where $\ln g(r)$ vs. $1/T$ plots are fairly straight lines (though with different slopes). 
The main variation of $g(r)$ with inverse temperature is seen near 
the first peak of the $g(r)$, {\it i.e.}, near $r=1.0$. Fig.~\ref{gr-invT}(b) exhibits the 
$1/T$-dependence of $g(r)$ at $r=1.0$. 
{It shows a non-monotonic behavior with maximum near $T=3.0$.} 
On either side of the peak the plot is a straight line, and thus the van't Hoffian assumption 
holds good in two temperature ranges, one at low $T$ and another at high $T$. 

{ This peak position should not  taken to be an isosbestic point. Isosbestic points have been observed in oxygen-oxygen radial distribution 
	function $g_{OO}(r)$ in water\cite{skinner,pathak,bosio} where 
	${\partial g(r)}/{\partial \beta}=0$. In the present case ${\partial g(r)}/{\partial \beta}$ at $r=1.0$ is zero due to the presence of a peak unlike 
	isosbestic points which are temperature independent. The first isosbestic point for the LJ system  at density $\rho=0.80$ is $1.33$ (not shown). 
	}

\subsection{{Application of Piskulich-Thompson theory }}

\begin{figure*}
	\includegraphics[width=0.40\textwidth,angle=0]{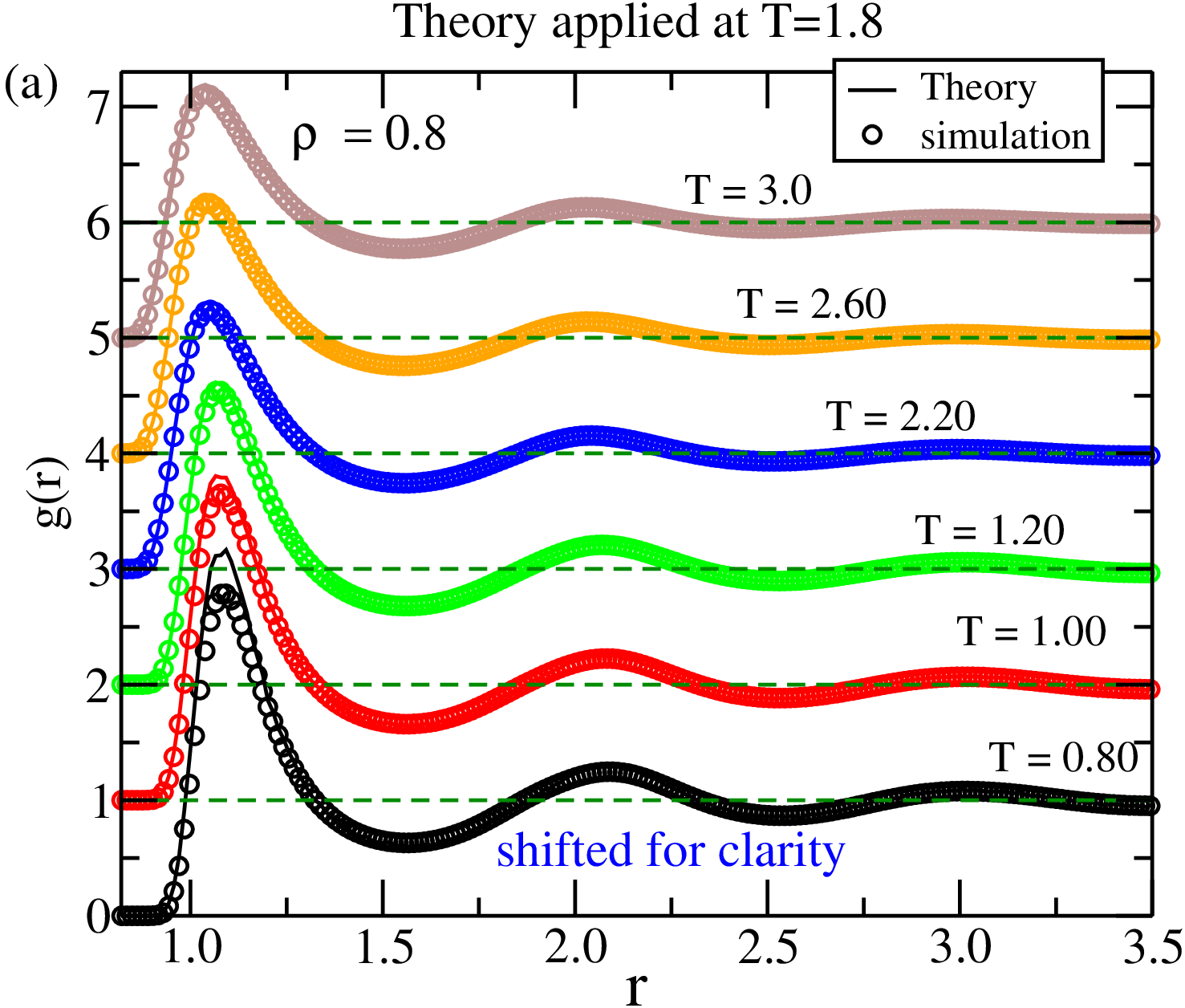}\hspace{3mm}\includegraphics[width=0.40\textwidth,angle=0]{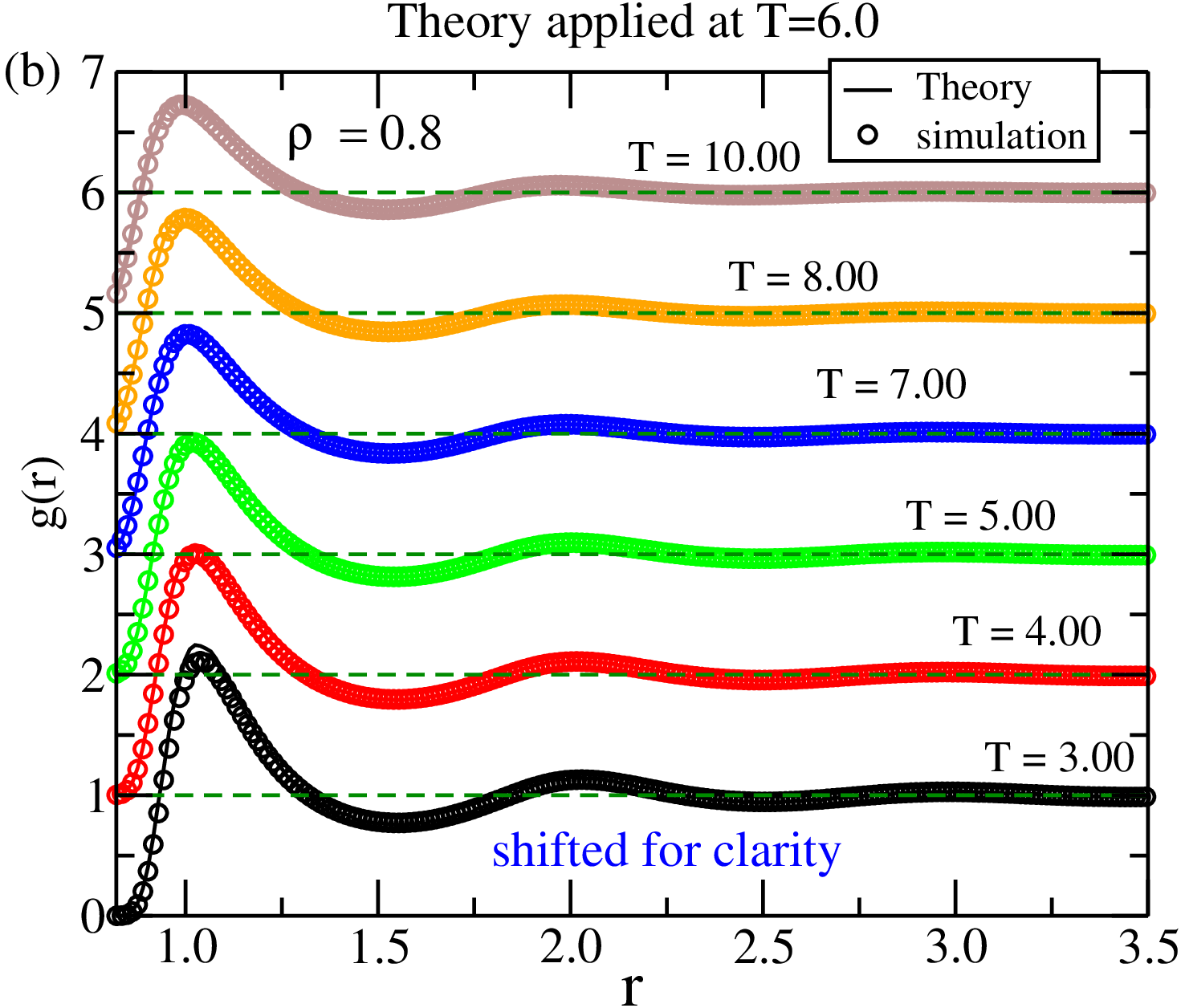}\vspace{11mm}
	\includegraphics[width=0.40\textwidth,angle=0]{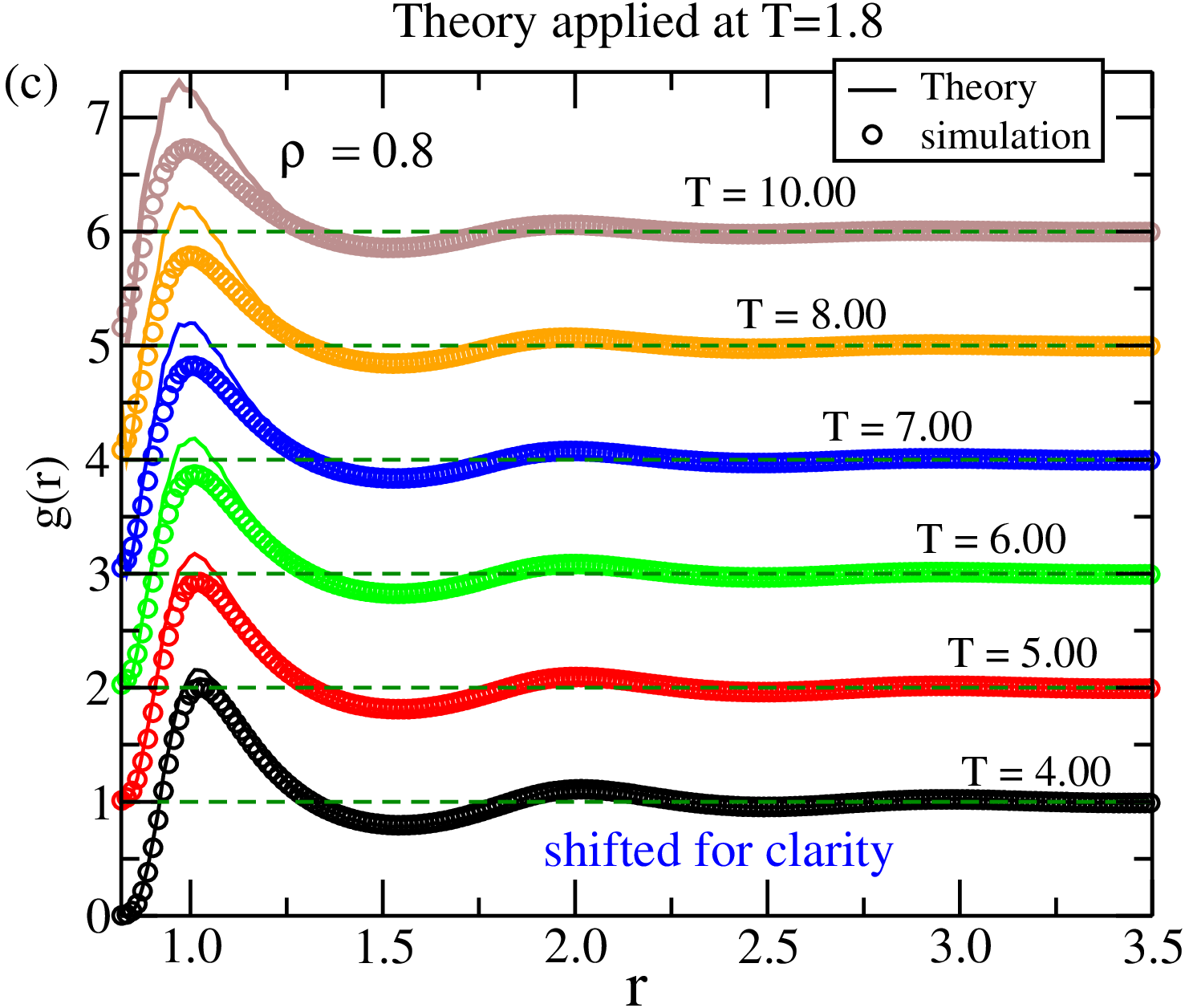}\hspace{3mm}\includegraphics[width=0.40\textwidth,angle=0]{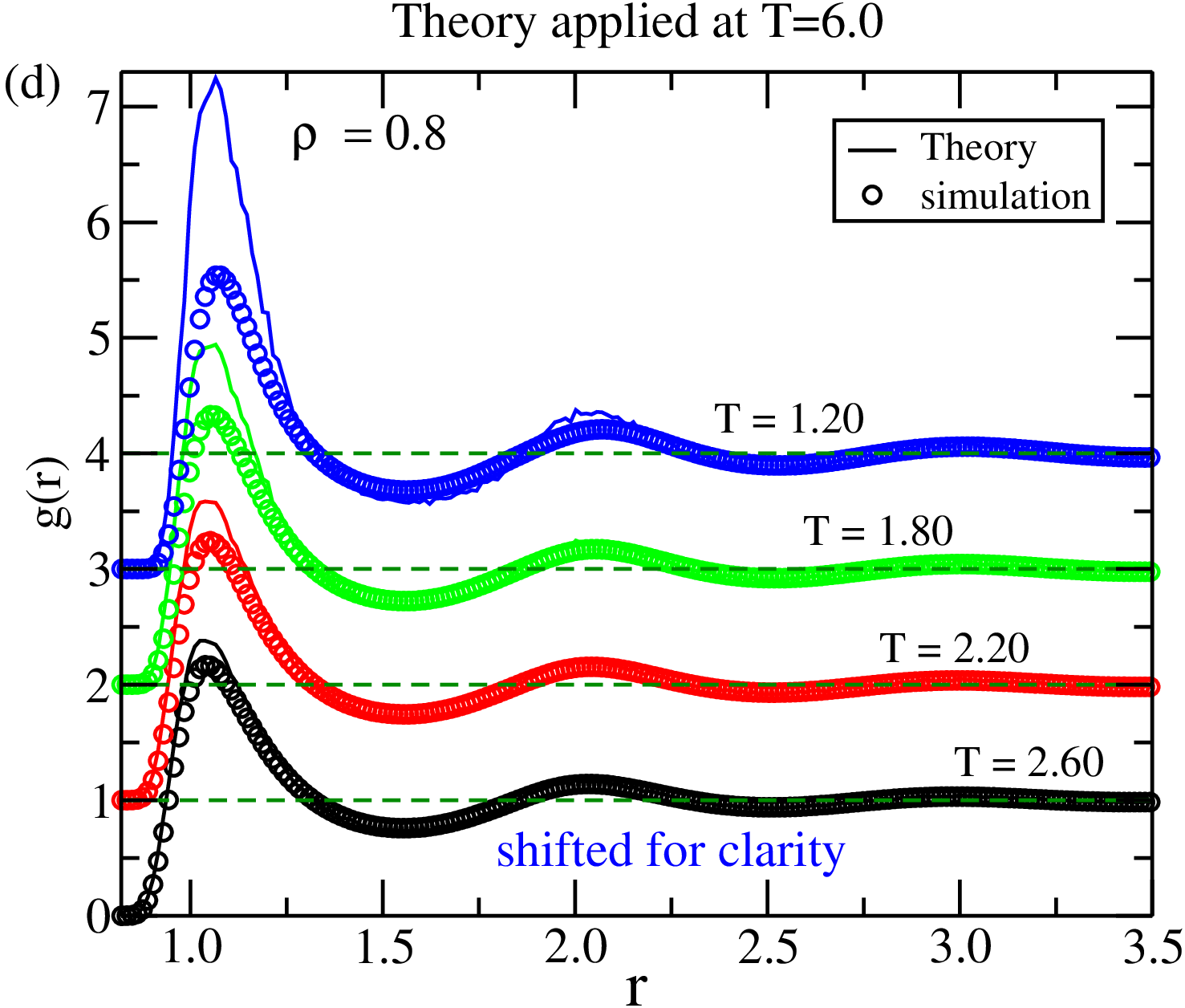}\vspace{0mm}
	{
	\caption{A comparison of $g(r)$ estimated at various temperatures $T$ by employing the Piskulich-Thompson theory at a reference temperature $T_r$ with that from simulations.
	(a) Both $T_r=1.8$ and various temperatures $T$ are on the low-temperature side of the peak of $g(r=1.0)$ vs. $1/T$ in Fig.~\ref{gr-invT}(b).
	(b) Both $T_r=6.0$ and various temperatures $T$ are on the high-temperature side of the peak.
	(c) $T_r=1.8$ and various temperatures $T$ are on the low- and high-temperature sides of the peak, respectively.
	(d) $T_r=6.0$ and various temperatures $T$ are on the high- and low-temperature sides of the peak, respectively.
	The density is $\rho=0.80$.}
\label{gr-compare-rho0.8}
	}
\end{figure*}

We now apply the Piskulich-Thompson theory at two temperatures, one on each side of the peak in Fig.~\ref{gr-invT}(b) where the 
van't Hoffian assumption is approximately valid, in order to determine $g(r)$ at other temperatures on that side.
Fig.~\ref{gr-compare-rho0.8}(a) exhibits $g(r)$ at $T=0.80, 1.00, 1.20, 2.20, 2.60$, and $3.00$ obtained 
by applying Piskulich-Thompson theory at reference temperature $T=1.8$ and density $\rho=0.80$. The radial distribution 
functions determined by employing Piskulich-Thompson theory (lines) have been compared with that obtained from 
MD simulation (symbols). They are in good agreement. The data have been shifted upward for clarity. 
Fig.~\ref{gr-compare-rho0.8}(b) shows the comparison of $g(r)$ at $T=3.0, 4.0, 5.0, 7.0, 8.0$, and $10.0$ determined from 
Piskulich-Thompson theory applied at $T=6.0$ (lines) with that obtained from direct MD simulations (symbols). 
The good agreement of the two $g(r)$ illustrates the following points: (i) the van't Hoffian assumption is approximately valid in two separate 
temperature ranges at each side of the $g(r=1.0)$ peak in Fig.~\ref{gr-invT}(b); 
(ii) the Piskulich-Thompson theory works for the LJ system in two temperature ranges. 

{
	Fig.~\ref{gr-compare-rho0.8}(c) shows the comparison between theoretical and simulation $g(r)$ at temperatures $T=4.0, 5.0, 6.0, 7.0, 8.0$, and $10.0$, which are
	on the high-temperature side of the peak in Fig.~\ref{gr-invT}(b). The Piskulich-Thompson theory has been applied at $T=1.8$, which is on the low-temperature 
	side of the peak in Fig.~\ref{gr-invT}(b). With the increase of temperature, $g(r)$ obtained from theory deviates from the simulation one. A 
	comparison of $g(r)$ for temperatures on the low-temperature side of the peak in Fig.~\ref{gr-invT}(b) are shown in Fig.~\ref{gr-compare-rho0.8}(d), where the theory has been applied at
	$T=6.0$, which is on the high-temperature side of the peak in Fig.~\ref{gr-invT}(b). The two $g(r)$ again show disagreement, which worsens with the temperature moving away from
	$T=3.0$, where $g(r)$ vs. $1/T$ shows a peak. When the theory is applied at the peak position ($T=3.0$), the disagreement between theoretical and simulation $g(r)$ is seen 
	on both sides of $g(r)$ vs. $1/T$ peak (see Appendix).
	Fig.~\ref{phase_diag} shows the van't Hoff demarcation line in the temperature-density phase-diagram of the LJ system. The van't Hoff demarcation line has been estimated from the peak positions 
	of $g(r)$ vs. $1/T$ for different isochores (see Appendix). The $g(r)$ vs. $1/T$ shows a peak for several values of $r$. We have considered $r$ satisfying $\rho^{1/3}r=0.8^{1/3}$
	in order to be close to the first peak position of the radial distribution function.
	The van't Hoff demarcation line increases with density.
	Fig.~\ref{phase_diag} also shows the melting line\cite{Lorenzo2016}, freezing line\cite{Lorenzo2016}, and liquid-gas coexistence curve\cite{Heyes2015}. The two isomorphs 
        shown are discussed in section~\ref{pt-iso-theory}.
	Though the van’t demarcation line is well above the critical temperature, it is still much lower than the Frenkel line\cite{Yoon}.
	At density $\rho=0.80$, temperatures of the Frenkel and van't Hoff demarcation lines are around $T=14.0$\cite{Yoon} and $T=3.0$, respectively.
}

\begin{figure}[t]
	\includegraphics[width=0.40\textwidth,angle=0]{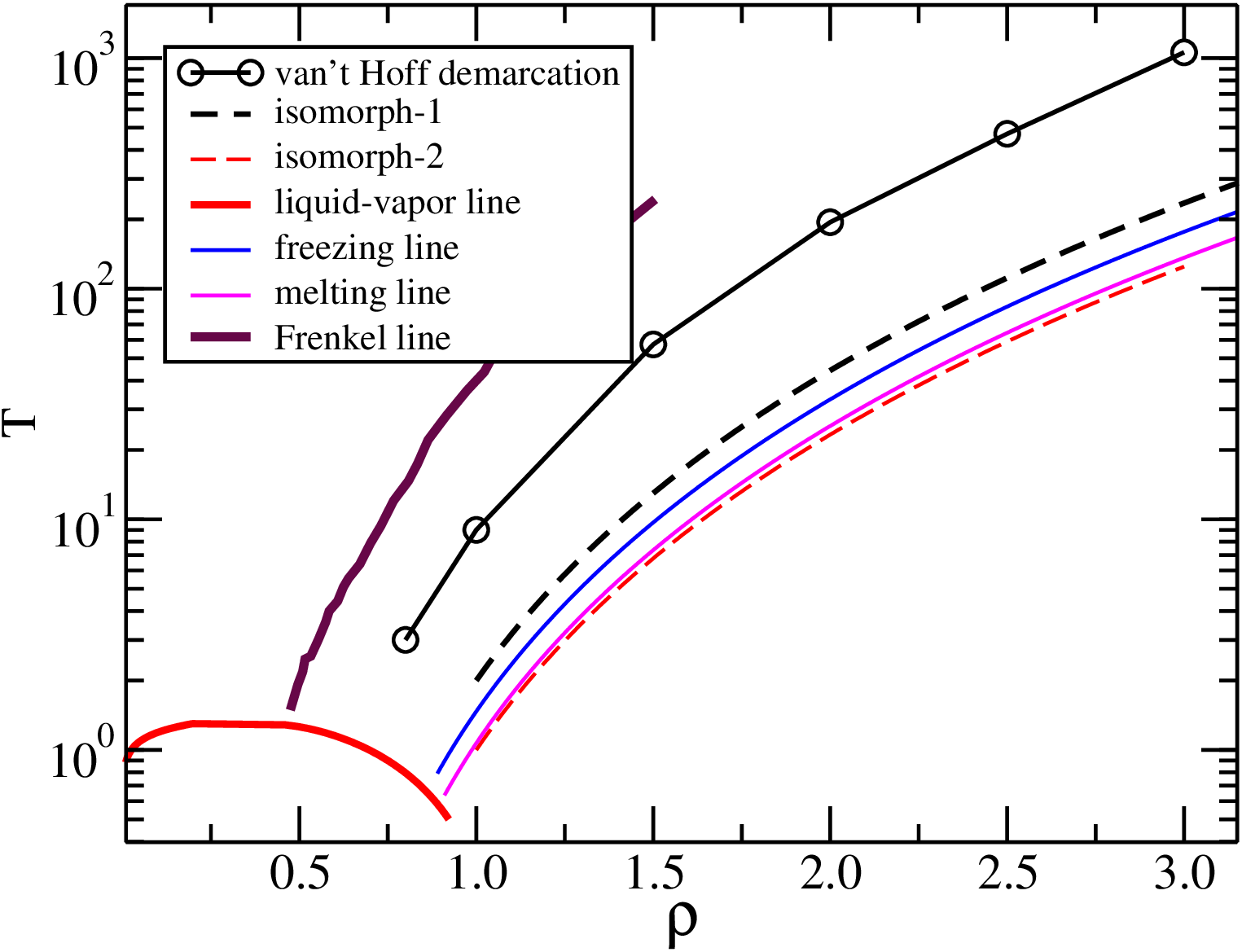}\vspace{0mm}
	{
		\caption{The van't Hoff demarcation line, obtained from the peak positions of $g(r=(0.80/\rho)^{1/3})$ vs. $1/T$ for various isochores, 
		has been shown in the LJ phase-diagram. The freezing line\cite{Lorenzo2016}, melting line\cite{Lorenzo2016}, liquid-gas coexistence curve\cite{Heyes2015} 
		and Frenkel line\cite{Yoon} have also been shown.}
	\label{phase_diag}
	}
\end{figure}

\section{The structure at an arbitrary temperature along an isochore estimated from  two radial distribution functions} \label{no-sim}
Equation \eqref{grT-mine} can be re-written as
{
\begin{eqnarray}
		U(r) &=& \dfrac{k_B T_0 T}{T-T_0} \ln \left[ \dfrac{g(r;\beta)}{g(r;\beta_0)} \right],\label{Ur2}
\end{eqnarray}
}
where $\beta_0 = \dfrac{1}{k_BT_0}$. 
{
It is easy to show that the first-order Taylor expansion of $g(r;\beta)$ around $g(r;\beta_0)$ reduces Eq.\eqref{Ur2} to Eq.\eqref{U} in the limit of $T \rightarrow T_0$.
	}
Thus $U(r)$ can be evaluated from $g(r)$ at two different temperatures at a given density.
The procedure for obtaining $g(r)$, whether in experiments or simulations, is irrelevant.
These two temperatures can be anywhere on the isochore in question, as long as the van't Hoffian assumption 
is valid. However, one has no prior knowledge of the temperature range where van't Hoffian assumption is valid 
for the system under consideration. It is, therefore, intuitive to consider two temperatures that are not far away from each other, and hence 
the van't Hoff's assumption is valid  at least in that small temperature range. 
Once the $ U(r)$ is determined, the $g(r)$ of 
liquids can be calculated at any temperature along the isochore from equation \eqref{grT-mine} without 
performing further simulation (or conducting more experiments). This is quite useful for a liquid with 
unknown interparticle interaction, and hence in this sense, the method is superior to standard liquid state theory, which 
requires the knowledge of the interactions between the particles.

\begin{figure}[t]
	\includegraphics[width=0.40\textwidth,angle=0]{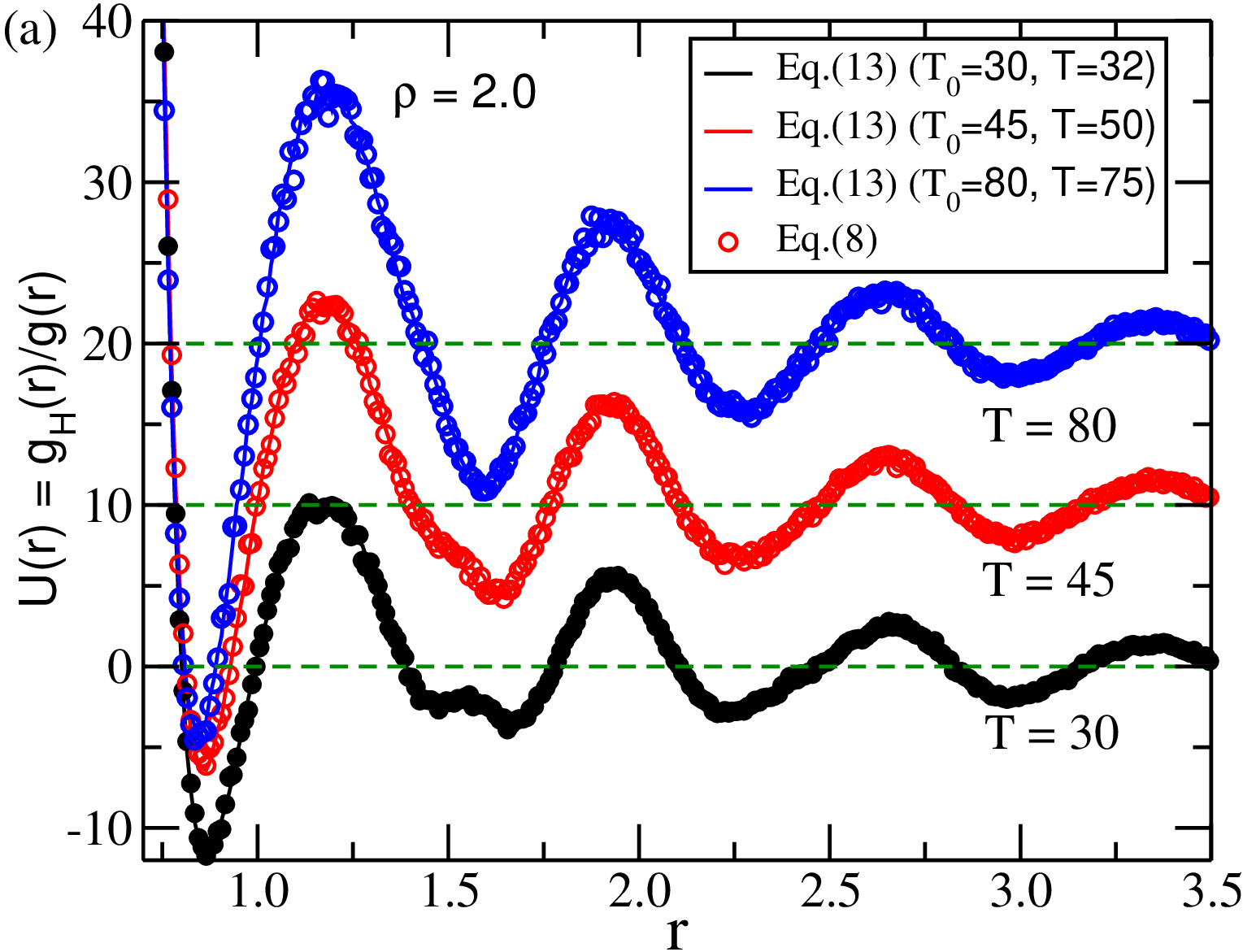}\vspace{10mm}
	\includegraphics[width=0.40\textwidth,angle=0]{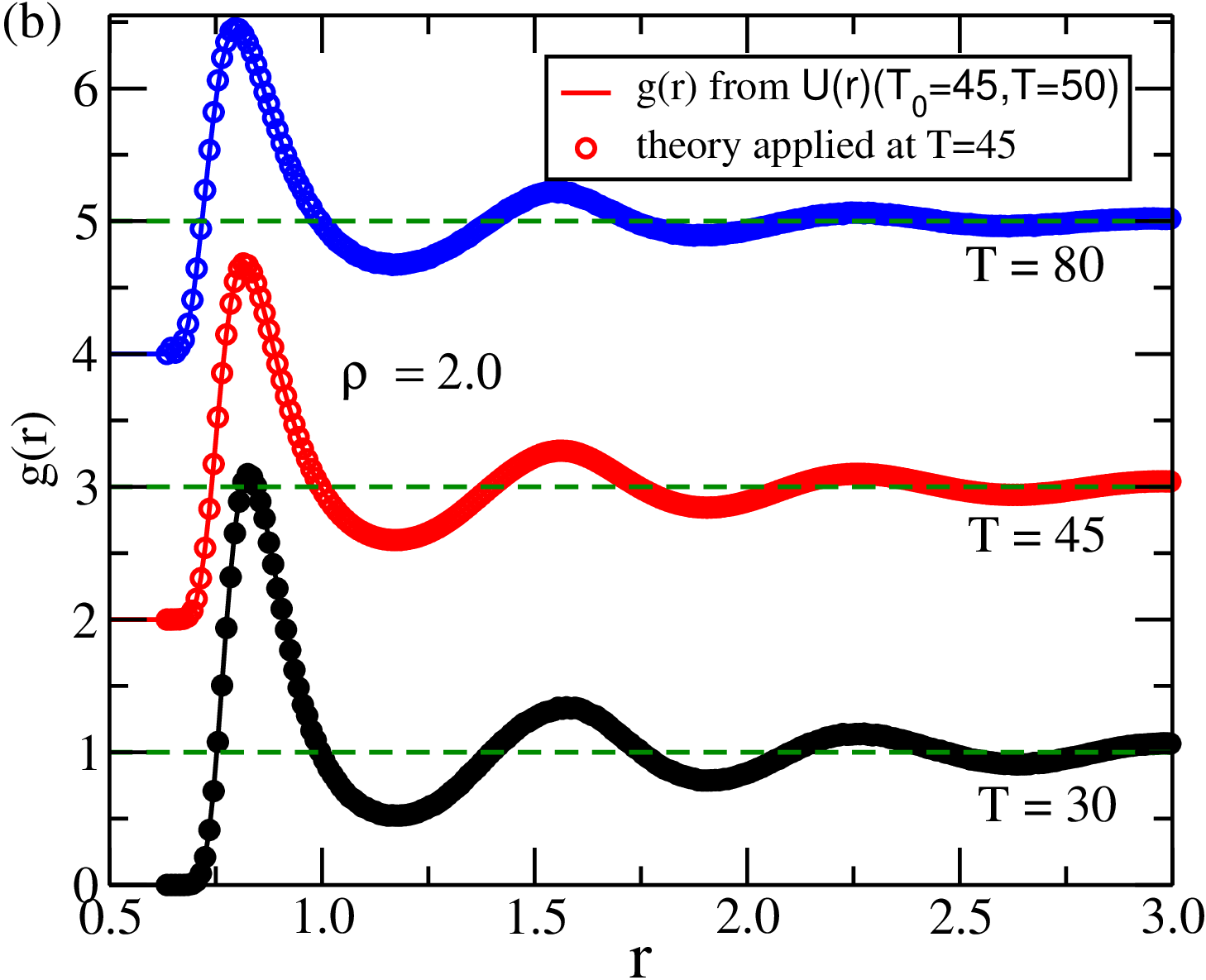}\vspace{0mm}
	{
		\caption{(a) The comparison of energetic force $U(r)$ for the LJ system at density $\rho=2.0$ and $T=30$, $T=45$ and $T=80$ obtained from fitting to Eq.\eqref{Ur2} and Eq.\eqref{U}. The data have been shifted upward for clarity by $10$ units. (b) Comparison of $g(r)$ obtained from Piskulich-Thompson theory with one obtained from $U(r)$ using Eq.\eqref{Ur2} instead of from simulation data. The data has been shifted upward by $2$ units for clarity.
	}
	\label{gr-from-Ur}
	}
\end{figure}

{
	Fig.~\ref{gr-from-Ur}(a) shows the comparison of $ U(r)$ obtained by using Eq.\eqref{Ur2} and Eq.\eqref{U} 
	at $T=30$, $T=45$, and $T=80$ at density $\rho=2.0$. 
}
The $ U(r)$ at $T=30$ has been evaluated using $g(r)$ at temperatures $T=30$ and $T=32$. 
Similarly, the function $U(r)$ at $T=45$ and $T=80$ has been obtained from $g(r)$ at $T=45$ \& $T=50$ 
and $T=80$ \& $T=75$, respectively. The $ U(r)$ at these three temperatures obtained by using Eq.\eqref{Ur2} (lines) are 
in good agreement with that obtained from MD simulations (symbols) directly. 
{
	In MD simulations Eq.\eqref{U} is employed to evaluate $U(r)$.
	Thus, unlike the Piskulich-Thompson theory, one does not need the fluctuation of potential energy $\Delta {\cal U}$ to evaluate $U(r)$.
	Fig.~\ref{gr-from-Ur}(b) shows a comparison of $g(r)$ obtained from $U(r)$ evaluated by employing Eq.\eqref{Ur2}($T_0$=45,T=50) with the 
	one determined by employing the Piskulich-Thompson theory at $T=45$ (symbols) at temperatures $T=30$, $45$, and $80$. 
	}
They are in good agreement with one another. 
{Filled symbols for the temperature $T=30$ indicates that the liquid is supercooled. The theory works alike for normal and supercooled liquids.}	
This method opens up the possibility of predicting the $g(r)$ along an isochore of a liquid for which the interparticle interactions are 
unknown if its $g(r)$ at two nearby temperatures are available, 
say, from light scattering experiments. This method is robust for the temperature range where the van't Hoffian 
assumption is valid.

\section{Piskulich-Thompson+Isomorph theory}\label{pt-iso-theory}
So far we have discussed the determination of $g(r)$ along an isochore either (i) directly using Piskulich-Thompson
theory  or (ii) by employing Eq.\eqref{grT-mine} and \eqref{Ur2} where only $g(r)$ at two temperatures is needed. 
Now we generalize this into a method to calculate  $g(r)$ at an arbitrary temperature in the $T-\rho$ phase-diagram
from just one simulation at the reference state point $(\rho_0,T_0)$. To achieve this we combine the Piskulich-Thompson 
theory with isomorph theory. First we determine the isomorph passing through the reference point $(\rho_0,T_0)$. 
The equation for an isomorph of the LJ system is given by\cite{Lorenzo,isomorph2012-jcp,isomorph2012-njp}
\begin{eqnarray}
	\dfrac{T(\rho )}{T_0} = \left(\dfrac{\gamma_0}{2} -1 \right) \left(\dfrac{\rho}{\rho_0} \right)^4 - \left(\dfrac{\gamma_0}{2} -2 \right) \left(\dfrac{\rho}{\rho_0} \right)^2
\end{eqnarray}
where the so-called density-scaling exponent $\gamma_0$ is calculated from equilibrium fluctuations at the reference state point by means of 
\begin{eqnarray}
	\gamma_0 = \dfrac{\left< \Delta {\cal U} \Delta W \right>} {<(\Delta {\cal U})^2>}. 
\end{eqnarray}
Here, $\Delta \cal U$ and $\Delta W$ are fluctuations in potential energy and virial. 

\begin{figure}[t]
	\includegraphics[width=0.40\textwidth,angle=0]{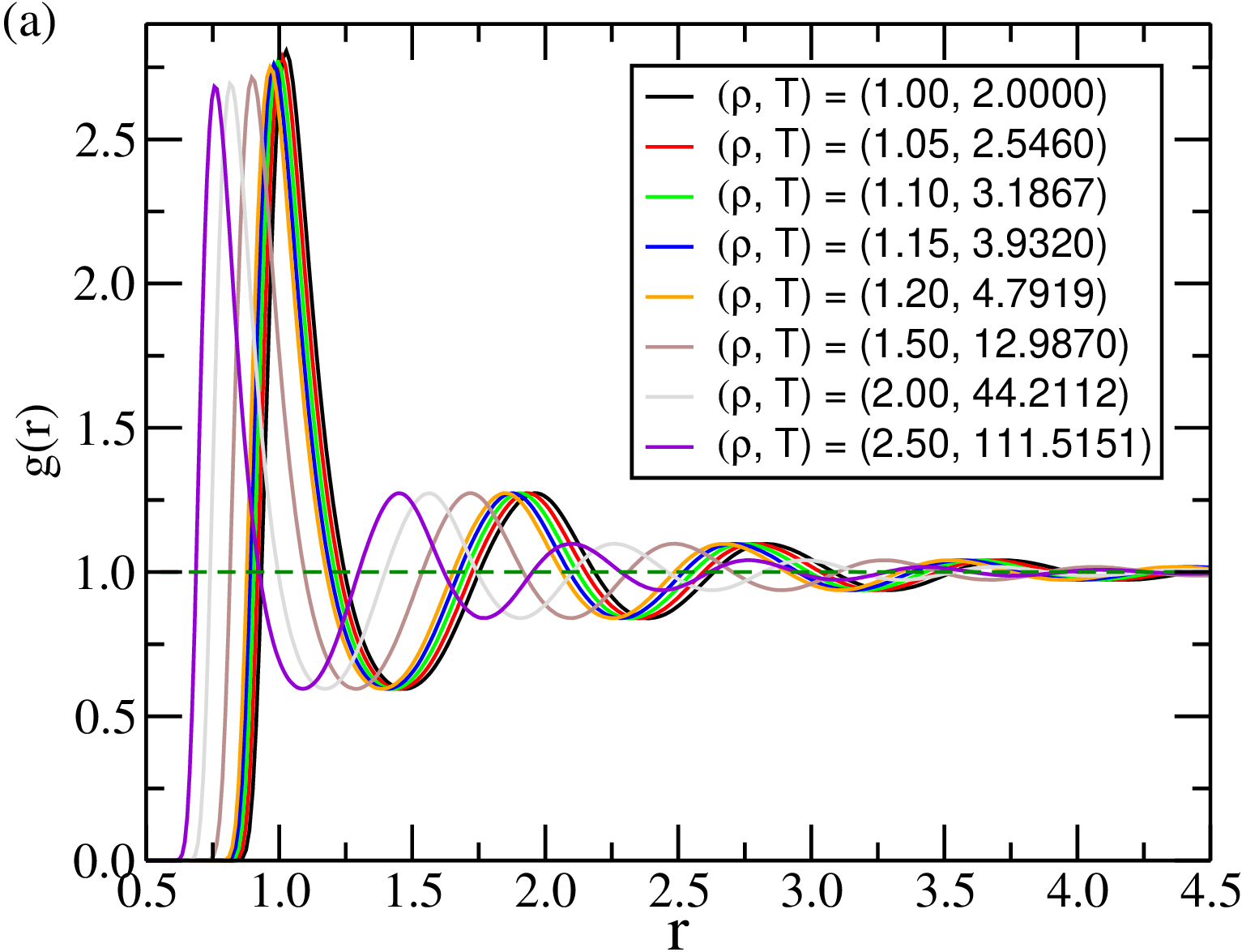}\vspace{10mm}
	\includegraphics[width=0.40\textwidth,angle=0]{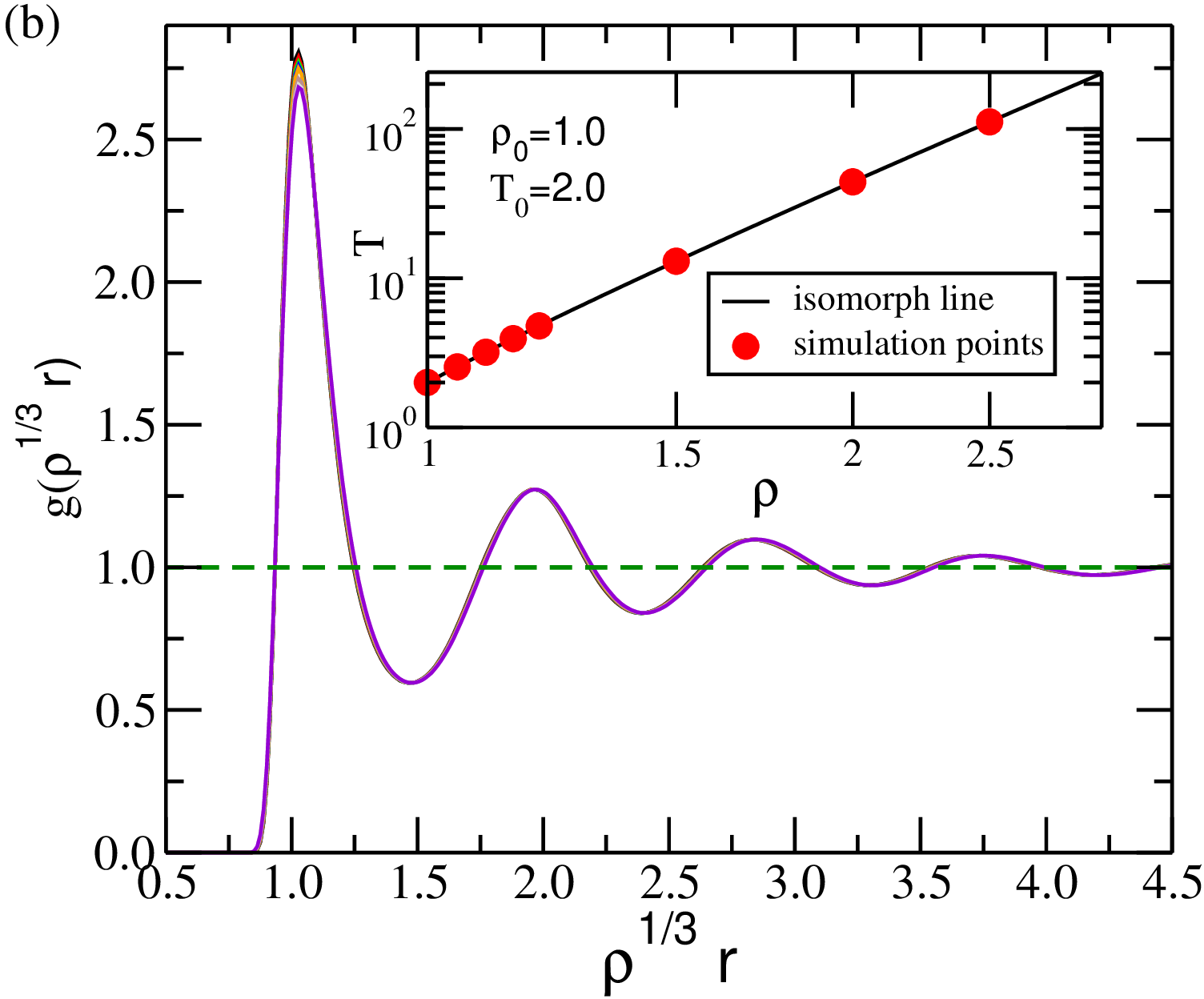}\vspace{0mm}
	\caption{(a) $g(r)$ at different state points on the isomorph; (b) same $g(r)$ in reduced units. The color scheme for both main panels is the same. Inset: isomorph and state points for which $g(r)$ have been shown here.}
\label{iso-gr}
\end{figure}

Fig.~\ref{iso-gr}(a) exhibits $g(r)$ at various state points along the isomorph starting from the 
{reference state point$(\rho_0,T_0)=(1,2)$}
for the LJ system. The inset of Fig.~\ref{iso-gr}(b) shows the isomorph (line) and the state points
(red symbols) where MD simulations have been performed. 
{The same isomorph is shown in Fig.~\ref{phase_diag} as {\it isomorph-1}, which is above the freezing line}.
The main panel of Fig.~\ref{iso-gr}(b)
shows the $g(r)$ in Fig.~\ref{iso-gr}(a) in reduced units. The color scheme for both figures are the same. 

As expected, $g(r)$ is invariant in  reduced units {i.e.} $g(\rho^{1/3}r)=contant$  along the isomorph
(Fig.~\ref{iso-gr}(b)). Hence, the $g(r)$ of the system at any point of the isomorph, say, 
($\rho_1,T_1)$,  can be obtained easily from the $g(r)$ of the reference point ($\rho_0,T_0)$. 
Thereafter, the Piskulich-Thompson theory can be employed along the isochore at $\rho_1$. In order to 
apply the Piskulich-Thompson theory we require the potential energy of all the configurations 
at ($\rho_1,T_1)$,  which is different from that at the reference state point $(\rho_0,T_0)$. However, 
the potential energy at $(\rho_1,T_1)$ can also be obtained by scaling the potential energy at the reference point ($\rho_0,T_0)$.

The scaled potential energy $\cal U$ of the LJ system at $(\rho_1,T_1)$ in terms of potential energy at the 
reference point$ (\rho_0,T_0)$ is given by\cite{Schroder2011}
\begin{eqnarray}
	{\cal U} = \tilde\rho^{m/3} {\cal U}_0^m + \tilde\rho^{n/3} {\cal U}_0^n,
\end{eqnarray}
where $\tilde\rho = \rho_1/\rho_0$ and ${\cal U}^k \equiv <\underset{i>j}\sum \nu_{ij}^{k}(r_{ij})  >$. The ${\cal U}^m$ 
and ${\cal U}^n$ are the repulsive and attractive parts of the LJ potential (and hence $m=12$ and $n=6$), respectively, implying that
\begin{eqnarray}
	{\cal U} = {\cal U}^m +  {\cal U}^n,
\end{eqnarray}
in which ${\cal U}_0^m$ and  ${\cal U}_0^n$ are the values of ${\cal U}^m$ and ${\cal U}^n$ at the reference 
point ($\rho_0,T_0)$ on the isomorph. The ${\cal U}^m$ and ${\cal U}^n$ are given by\cite{Schroder2011}
\begin{eqnarray}
	{\cal U}^m = \dfrac{3W  - m {\cal U}} {m-n}, \label{Um-scale}\\
	{\cal U}^n = \dfrac{-3W  + m {\cal U}} {m-n}. \label{Un-scale}
\end{eqnarray}
The above equation is based on the fact that $\dfrac{{\cal U}^k}{\rho^{k/3}}$=constant (ignoring the linear term  
in the shifted-force cutoff LJ potential, see Eq.\eqref{lj}).

\begin{figure}[t]
	\includegraphics[width=0.40\textwidth,angle=0]{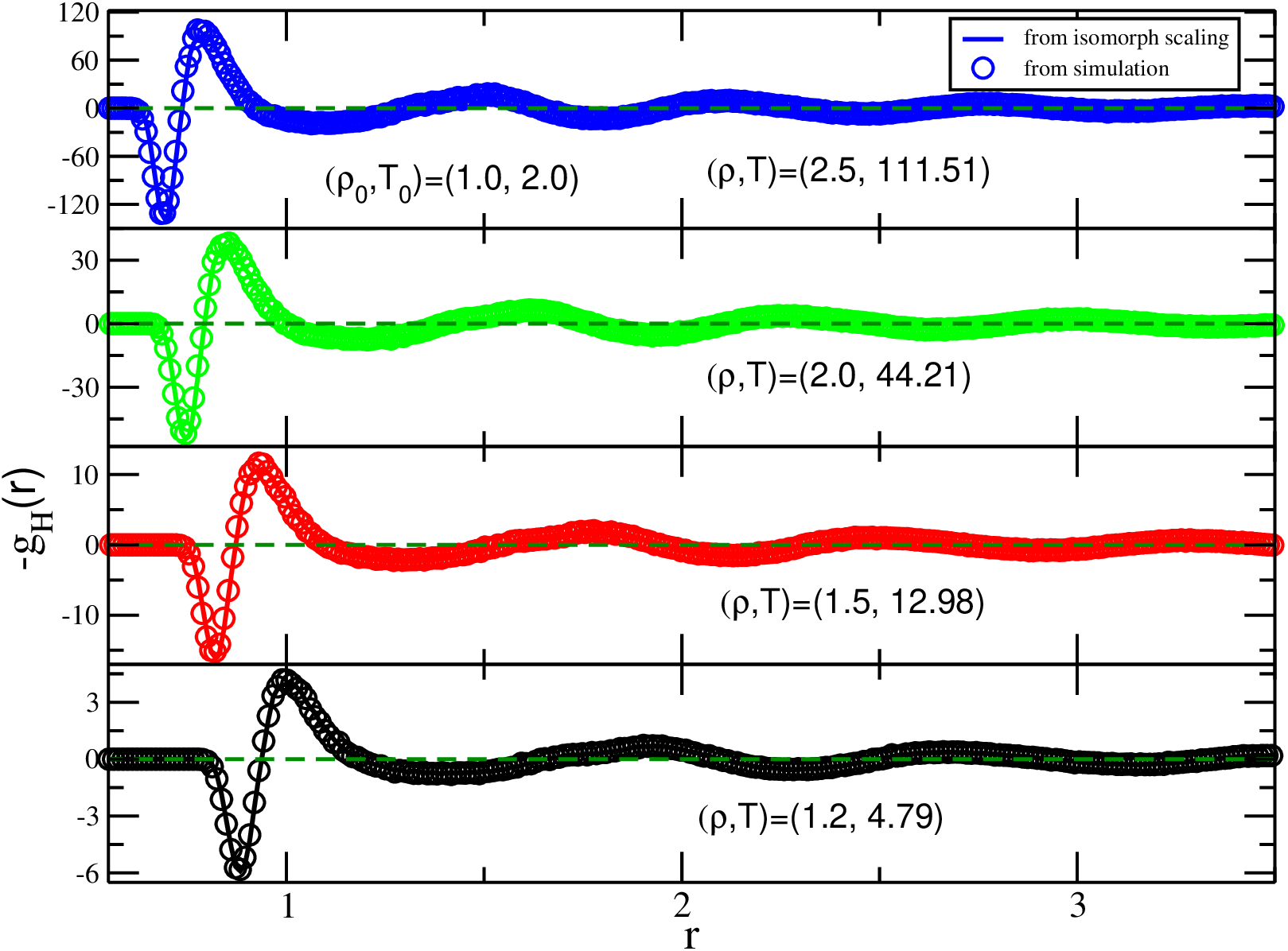}\vspace{1mm}
	{
	\caption{The comparison of $-g_H(r)$ obtained from isomorph scaling and direct simulation at different points on the isomorph.}
\label{gH-iso}
	}
\end{figure}

\begin{figure*}
	\includegraphics[width=0.40\textwidth,angle=0]{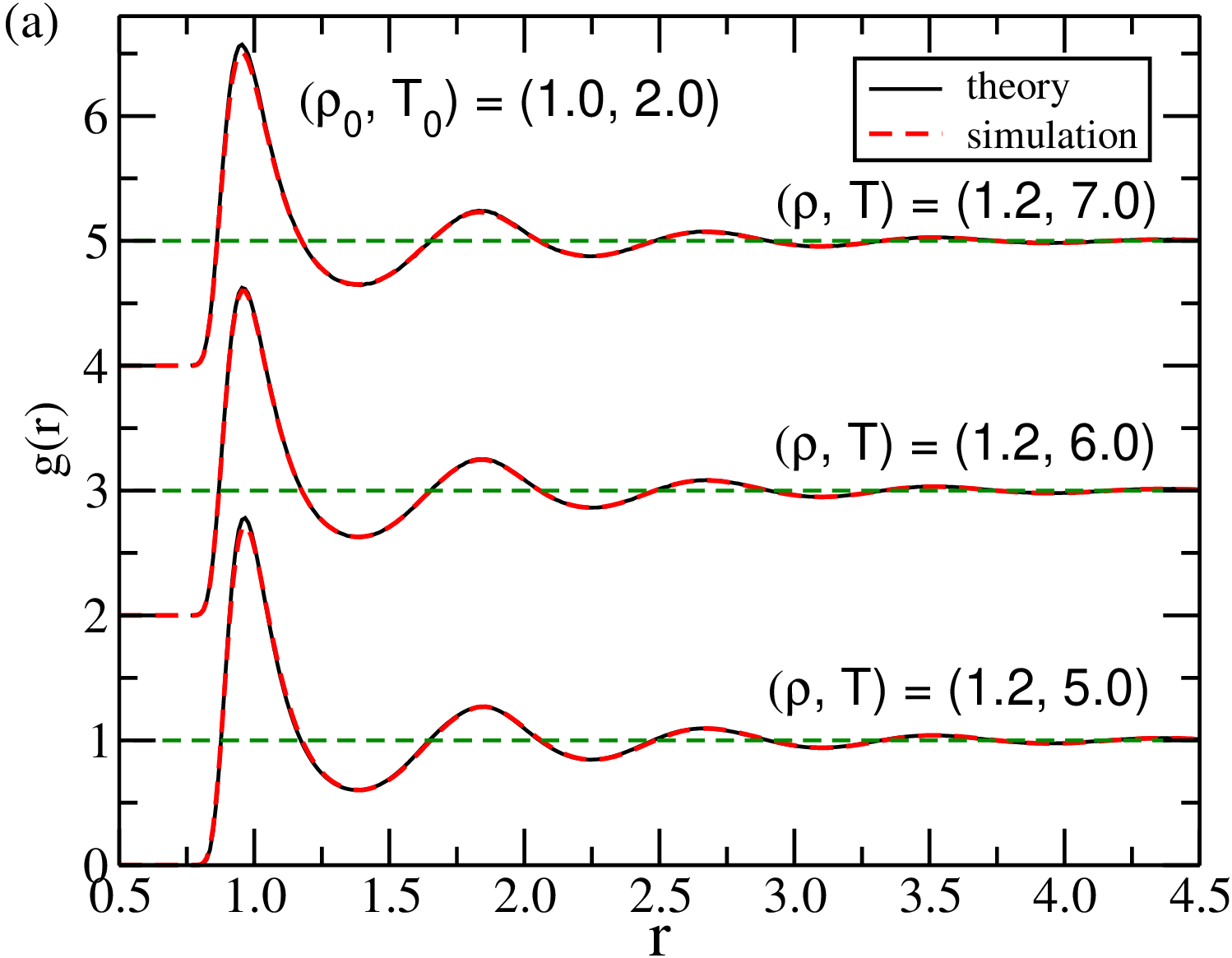}\hspace{3mm}\includegraphics[width=0.40\textwidth,angle=0]{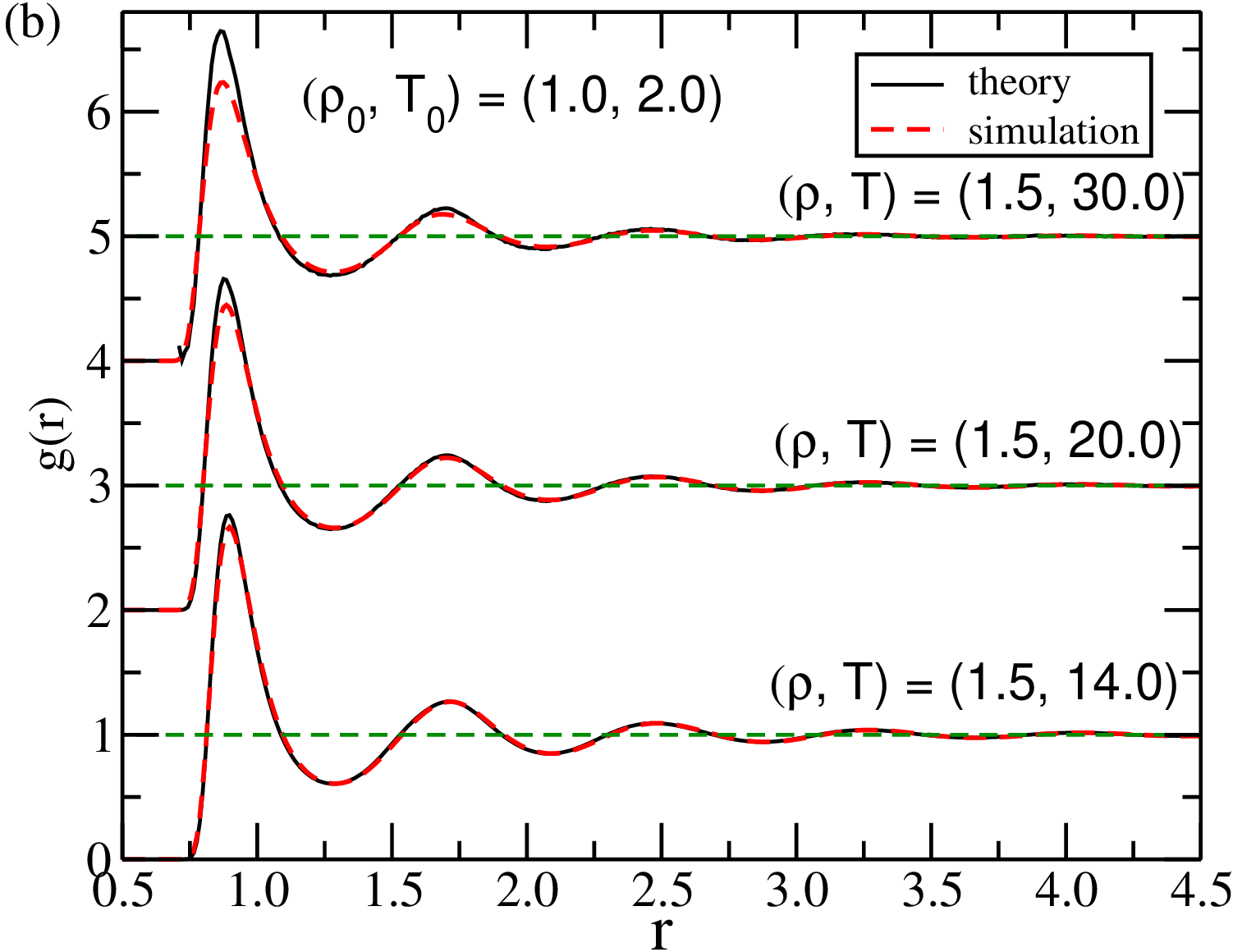}\vspace{9mm}
	\includegraphics[width=0.40\textwidth,angle=0]{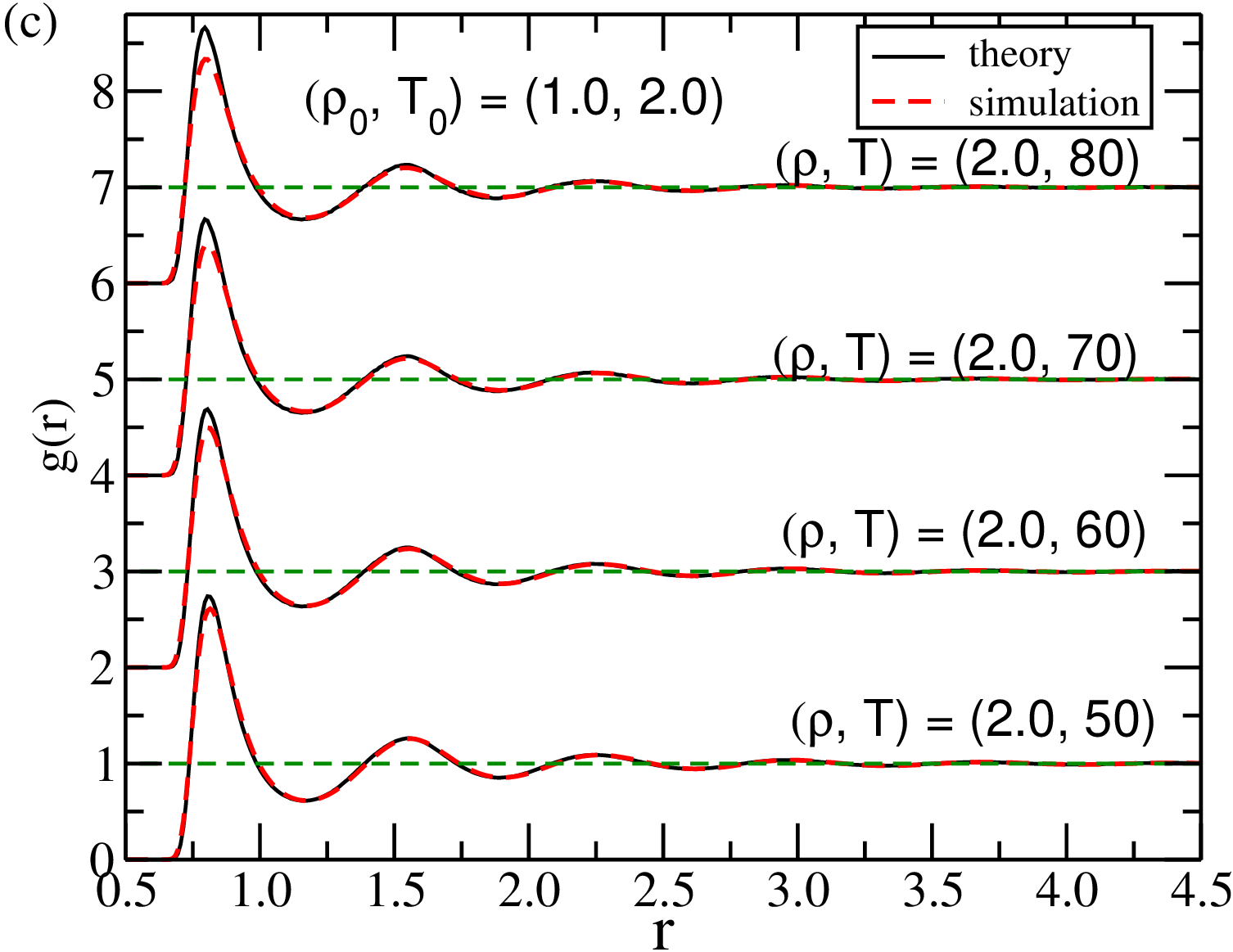}\hspace{3mm}\includegraphics[width=0.40\textwidth,angle=0]{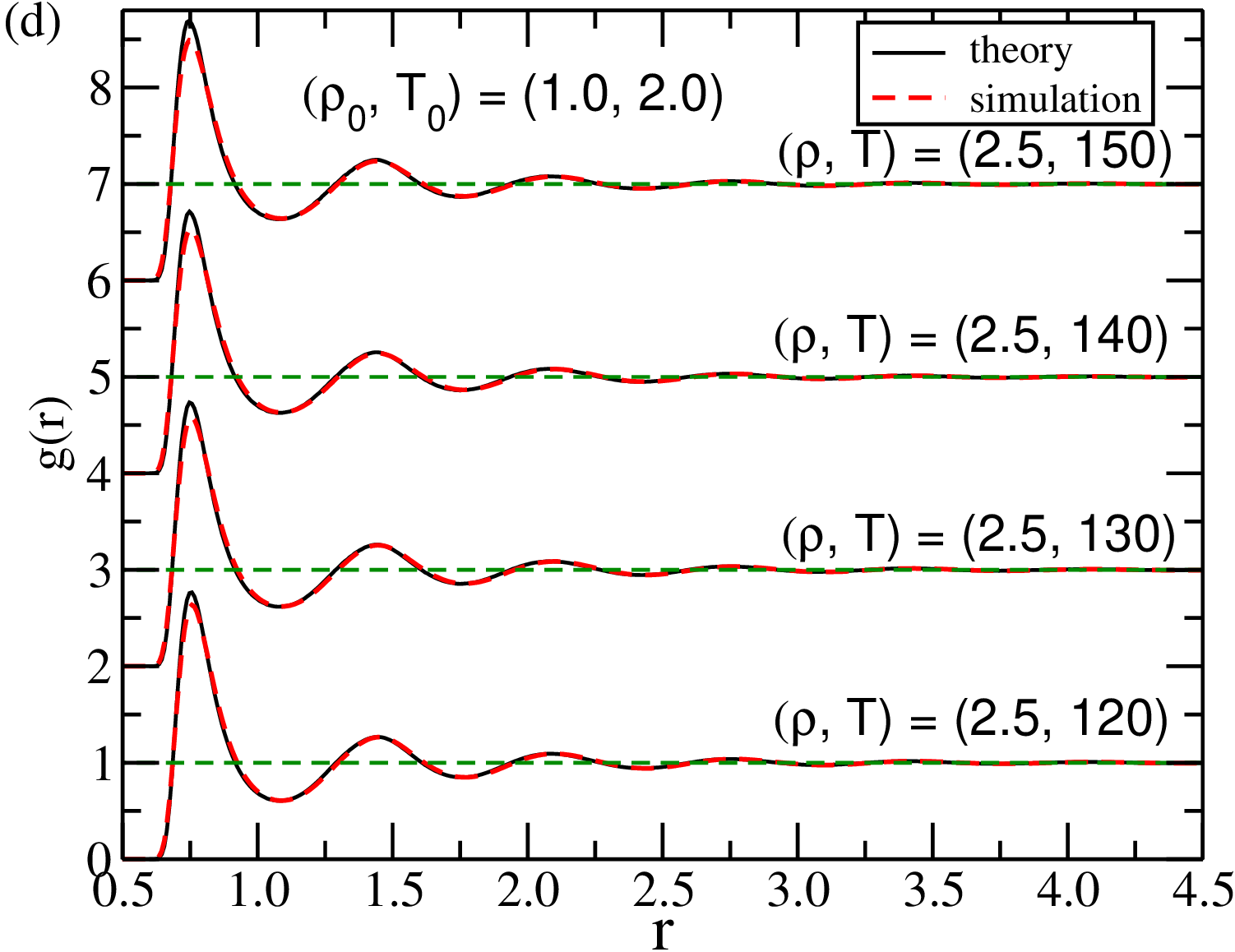}\vspace{0mm}
	{
	\caption{The comparison of $g(r)$ obtained by employing Piskulich-Thompson+isomorph theory with that from simulation at different state points of an isochore (a) $\rho=1.20$ (b) $\rho=1.50$ (c) $\rho=2.00$, and (d) $\rho=2.50$. The reference temperature and density for Piskulich-Thompson+isomorph theory is ($\rho_0=1.0, T_0=2.0$).}
\label{iso-rho1.1-2.5}
	}
\end{figure*}

{
Fig.~\ref{gH-iso} shows a comparison of $-g_H(r)$ obtained from isomorph scaling and direct simulation at state points ($\rho$=1.2, T=4.79), ($\rho$=1.5, T=12.98), 
($\rho$=2, T=44.21) and ($\rho$=2.5, T=111.51) on the isomorph. The $-g_H(r)$
}
obtained from isomorph scaling (lines) described above are in good agreement with those 
obtained from MD simulations (open symbols) at these state points. 
Now we have $g(r)$ as well as $-g_H(r)$ (so $U(r)$)  at the state point ($\rho_1,T_1)$, and therefore, the Piskulich-Thompson theory can be applied easily.
Figures ~\ref{iso-rho1.1-2.5}~(a)-(d) 
show comparisons of $g(r)$ obtained by employing Piskulich-Thompson+isomorph theory with those from simulations. 
They are in good agreement when $|T-T_1|$ is small. The discrepancy at large $|T-T_1|$ is associated with following two facts: 
(i) the LJ system obeys the van't Hoff's assumption to a good approximation only in a limited temperature range and 
(ii) the isomorph is incapable to capture the first peak correctly\cite{isomorph2009}
One can observe the discrepancy between isomorph theory and simulation in 
{$-g_H(r)$ near the first peak in Fig.~\ref{gH-iso}, as well.}
To summarize, for a van't Hoffian valid liquid, $g(r)$ can be calculated at any arbitrary state point in the liquid region of 
the temperature-density phase diagram from a single state point simulation.

\section{Discussions and Summary}\label{summary}

The van't Hoffian assumption is that the energetic and entropic forces are temperature independent.
We have shown that the LJ system disobeys the van't Hoff's assumption when viewed over the entire temperature range studied. 
Unlike other non-van't Hoff liquids\cite{Dorsey,Galaon}, the LJ system does not have any covalent bond. 
The fact that the van't Hoffian assumption breaks down might be
due to a different activation energy at low and high temperatures. While at very high temperature, the 
LJ system is governed by entropic forces and energy plays little role; at low temperature the energy 
dominates and some of the particles remain close to one another for quite a long time, behaving as a 
quasi-covalent bond. 

While the van't Hoffian assumption for the LJ system is not valid for the whole range of temperatures studied,
there are two distinct temperature ranges (see Fig.~\ref{gr-invT}(b)) where the van't Hoffian assumption applies approximately. This is validated by the excellent agreement of the $g(r)$ obtained from 
Piskulich-Thompson theory and simulation (see Fig.~\ref{gr-compare-rho0.8}(a)-(b)). 
The Piskulich-Thompson theory has been applied at one 
temperature in each low $T$-range ($T=1.8$) as well as high $T$-range ($T=6$) to determine $g(r)$ at other 
temperatures in that range (see Fig.\ref{gr-compare-rho0.8}(a)-(b)). For a van't Hoffian system,
just a single simulation is required to determine $g(r)$ at an arbitrary temperature at 
the same density employing Piskulich-Thompson theory. For a non-van't Hoffian system, such as the LJ, 
the range of temperatures, where Piskulich-Thompson theory can be applied to determine $g(r)$, 
is limited. Thus for such a system, one can only determine $g(r)$ at the temperatures in the vicinity 
of the state point where simulation data are available. Simulations of supercooled liquids are challenging due to their
long relaxation time and strong crystallization tendency. At such low temperatures, the van't Hoffian
assumption would probably be valid for all systems, and thus Piskulich-Thompson theory can be applied.
In such a scenario, this theory could be helpful.
{In the Appendix, we have shown that the theory indeed works equally well for supercooled liquids.}

We have shown that the energetic force $U(r)$ can be evaluated from a knowledge of $g(r)$ at two
temperatures, say $T_0$ and $T$, in the van't Hoffian region (see Eq.\eqref{Ur2}).
This method is particularly useful when the interatomic/intermolecular interactions are not known,
forbidding computer simulations, or when simulations / experiments are extremely challenging. Since many systems may 
have multiple temperature ranges with valid van't Hoffian assumption similar to the the LJ system, it is 
imperative to consider $g(r)$ at two very close temperatures, where the van't Hoff's assumption is bound to be valid.
Once $U(r)$ is determined, the $g(r)$ can be predicted at various temperatures along the isochore. 


The Piskulich-Thompson theory works only along an isochore. We have extended this theory to calculate 
the $g(r)$ at an arbitrary state point of the liquid region of the phase diagram from a single simulation at a 
reference state point $(\rho_0,T_0)$. For this, we have combined the Piskulich-Thompson approach with the isomorph theory. The 
structure of a liquid along an isomorph is invariant in reduced units. It should be noted that not all systems have 
isomorphs (are R-simple) \cite{non-isomorph} -- water is a striking counter example -- and the current theory is of course 
limited to R-simple liquids. The isomorph theory is not valid in the gaseous region of the LJ system as well\cite{isomorph2009,Paddy2013}, 
and hence, this theory can not be applied to determine the radial distribution function in the 
low-density region of the LJ phase diagram.

In order to calculate $g(r)$ at an arbitrary state point $(\rho,T)$, we first calculate $g(r)$ at $(\rho,T_{iso})$, 
where $T_{iso}$ is on the isomorph. We here need to scale the potential energy from the reference point to the 
$(\rho,T_{iso})$ as well. For the LJ system, this is done as shown in Eq. (\ref{Um-scale}-\ref{Un-scale}) following Ref. \onlinecite{Schroder2011}.
This expression is system dependent, and one needs to find the expression for other potentials as per the isomorph theory described in
the Ref.~\cite{Schroder2011}. Thus $ U(r) = g_H(r)/g(r)$ is known at $(\rho,T_{iso})$. 
{
Thereafter, the Piskulich-Thompson theory is employed along the isochore $\rho$ to calculate $g(r)$ at the designated state point $(\rho,T)$. 
}
Again, for a perfect van't Hoffian system $g(r)$ at every state point of the phase diagram (liquid region) can be obtained. 
On the other hand, if the system does not have a single temperature range where the van't Hoffian assumption is valid; 
this theory cannot be used to evaluate $g(r)$ in the whole liquid phase of the diagram from a single simulation.
But if the information of different temperature ranges and one simulation data in each temperature range are 
available, one can calculate $g(r)$ in the whole liquid phase part of the phase diagram for an R-simple system. 
We would again remind that not all systems have isomorphs\cite{non-isomorph,Dyre2014Review}. However, 
many systems are R-simple\cite{Dyre2014Review} and Piskulich-Thompson+isomorph theory should apply to any such system.

All thermodynamic quantities are related to the radial distribution function $g(r)$, and hence they can be evaluated in 
the liquid region of phase-diagram whenever this theory is applicable. But again, there are systems with 
three-body interactions such as silicon\cite{SW,saw2009prl,saw2011jcp} 
where this theory will not be applicable. As far as dynamics is concerned, the MCT 
requires the structure factor (which is Fourier transform of $g(r)$) and the interparticle 
interactions to provide the dynamics such as MSD and intermediate scattering 
function. Thus for an R-simple system, employing Piskulich-Thompson+isomorph theory along with MCT, 
one can calculate all thermodynamics as well as dynamical quantities.

In summary, we have shown that: (i) the LJ system disobeys the van't Hoff's assumption that 
the energetic and entropic forces are temperature independent. However, we have identified two temperature ranges 
in which the van't Hoffian assumption is valid to a good approximation. We validated this by comparing the $g(r)$ determined by 
employing Piskulich-Thompson theory with that obtained from the simulation with excellent agreement.  
(ii) one can obtain the energetic force term $ U(r)$ without any simulation, and only $g(r)$ at two temperatures in the
temperature range where van't Hoffian assumption is valid, is required. 
Then $g(r)$ along an isochore can be calculated from $U(r)$ at all temperatures where van't Hoff's
assumption is valid. (iii) the $g(r)$ can be determined at an arbitrary state point in the liquid region of the 
phase diagram for an isomorphic invariant (R-simple) liquid from just a single simulation by employing Piskulich-Thompson+isomorph theory. 

It would be interesting to investigate whether the van't Hoffian assumption is valid in the whole temperature range 
for other R-simple liquids, {\it e.g.}, inverse-power law, Yukawa potential or Morse potential systems. 

\section*{ACKNOWLEDGEMENTS}
This work was supported by the VILLUM Foundation's \textit{Matter} grant (No. 16515).
We thank Lorenzo Costigliola for the fruitful discussions.
\appendix
\makeatletter \renewcommand{\thefigure}{A\@arabic\c@figure} \makeatother
\setcounter {figure}{0}

\section{Piskulich-Thompson theory applied at the peak of $\ln g(r=1.0)$ vs. $1/T$}
\begin{figure}[t]
	\includegraphics[width=0.40\textwidth,angle=0]{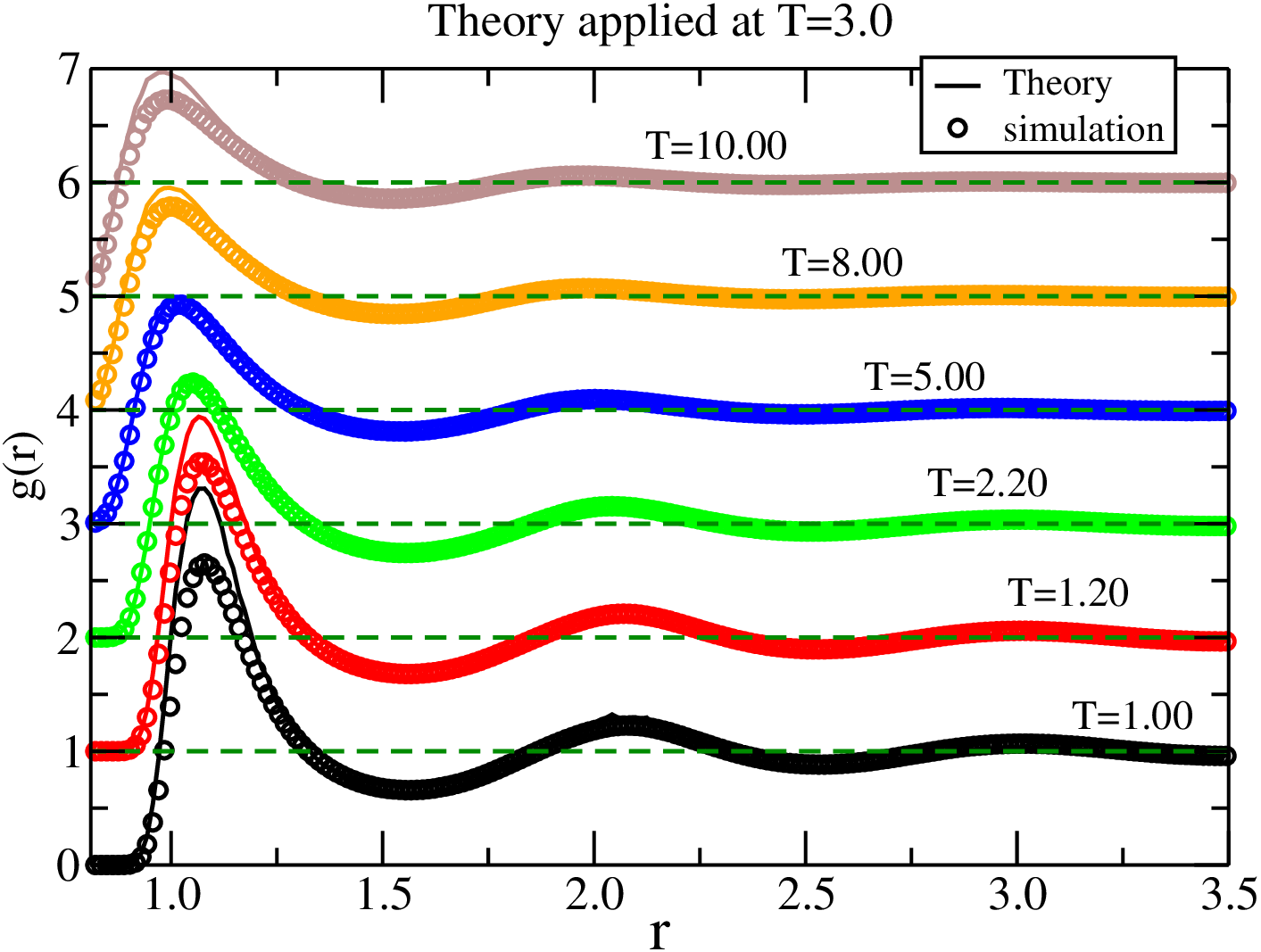}
	\caption{A comparison of $g(r)$ obtained from simulations and by employing the Piskulich-Thompson theory at the peak in Fig.~\ref{gr-invT}(b)(T=3.0), 
	at either side of the peak. The density is $\rho=0.80$.}
\label{theoryT3}
\end{figure}

Piskulich-Thompson theory is applied at the reference temperature $T=3.0$ for density $\rho=0.80$ where $\ln g(r=1.0)$ vs. $1/T$ exhibits a peak (see Fig.~\ref{gr-invT}(b)). 
From Fig.~\ref{theoryT3} it is evident that theory can predict the structure on either side of the peak only if the deviation from the reference temperature is small.
Theory fails to estimate $g(r)$ on either side of the peak when deviations from reference temperature are large.

\section{van't Hoff demarcation line}
\begin{figure}
	\vspace{5mm}
	\includegraphics[width=0.40\textwidth,angle=0]{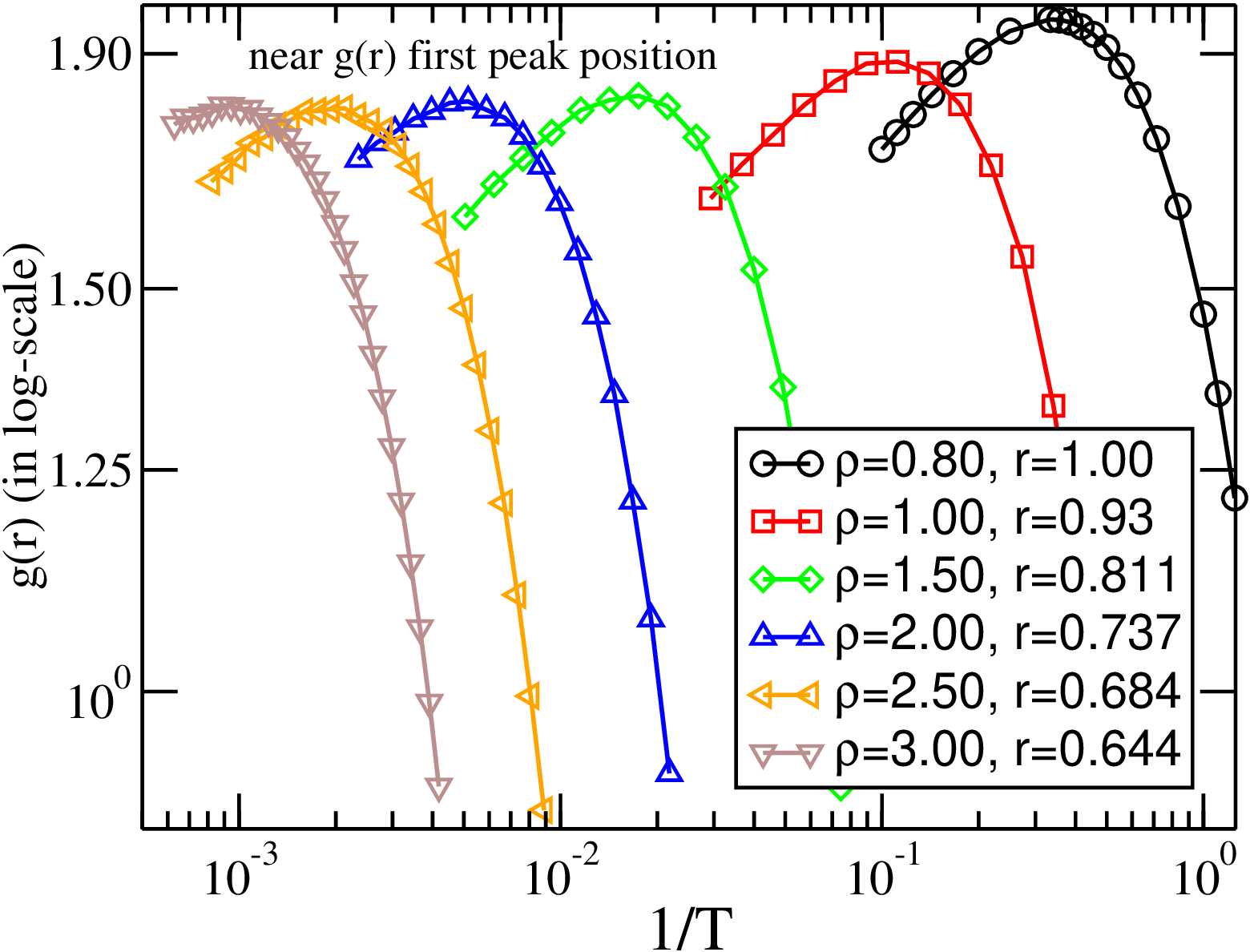}
	\caption{$T^{-1}$-dependence of $g(r)$ of the LJ system at various densities and $r$ satisfying $\rho^{1/3}r={0.80}^{1/3}$.}
\label{gr-invT_app}
\end{figure}
Fig.~\ref{gr-invT_app} exhibits the $1/T$-dependence of $g(r)$ for various isochores.
We have considered $r$ which satisfies $\rho^{1/3} r = {0.80}^{1/3}$.
The temperature, where $\ln g(r)$ vs. $1/T$ shows a peak, increases with  density.
These temperatures construct the van't Hoff demarcation line (see Fig.~\ref{phase_diag}).
The van't Hoff's assumption holds good on either side of the demarcation line in the phase diagram.
One can determine the $g(r)$ at an arbitrary state point on either side of the van't Hoff demarcation line from a single simulation on the same side.

\section{Piskulich-Thompson+isomorph theory applied along an isomorph in the supercooled regime}
\begin{figure}
	\vspace{11mm}
	\includegraphics[width=0.40\textwidth,angle=0]{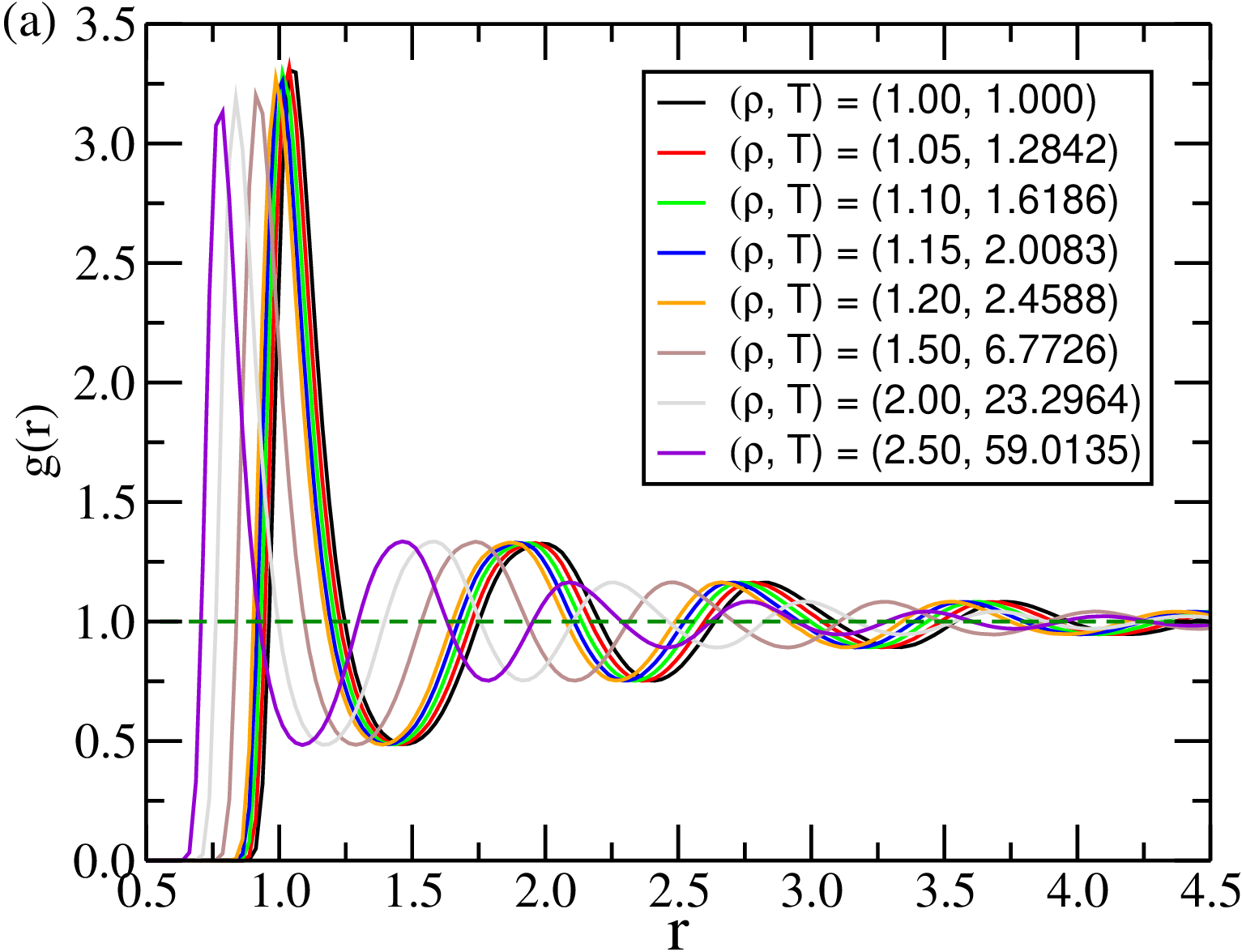}\vspace{10mm}
	\includegraphics[width=0.40\textwidth,angle=0]{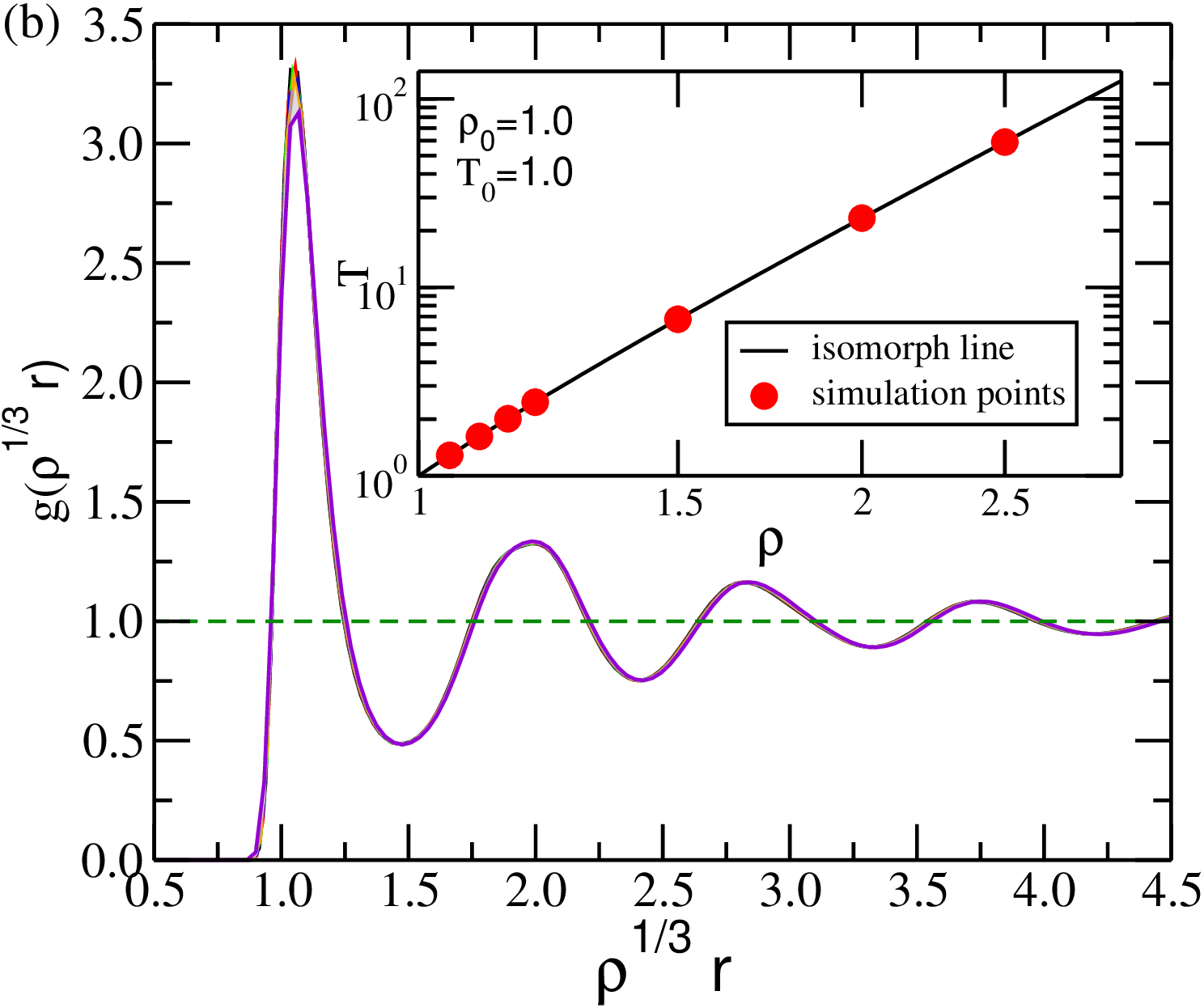}\vspace{0mm}
	\caption{(a) $g(r)$ at different state points on the isomorph starting from the reference state point $\rho_0=1.0$ and $T_0=1.0$ which is below the melting line; (b) same $g(r)$ in reduced units. The color scheme for both main panels is the same. Inset: isomorph and state points for which $g(r)$ have been shown here.}
\label{iso-gr_app}
\end{figure}

In the main text, we have shown that the Piskulich-Thompson+isomorph theory works well for normal LJ liquids.
However, this theory works equally well in the supercooled regime of the LJ phase diagram. For supercooled liquids, 
the reference state point is $(\rho_0,T_0)=(1,1)$ and a supercooled isomorph is shown by $isomorph-2$ in the Fig.~\ref{phase_diag}.
Fig.~\ref{iso-gr_app}(a) shows the $g(r)$ at different state points along the isomorph in the supercooled regime, starting from the reference state point $(\rho_0,T_0)=(1,1)$.
The inset shows the  state points where MD simulations are performed. Fig.~\ref{iso-gr_app}(b) shows that the $g(r)$ in the reduced units, {\it i.e.} $g(\rho^{1/3}r)$ is isomorph invariant.

\begin{figure}
	\includegraphics[width=0.40\textwidth,angle=0]{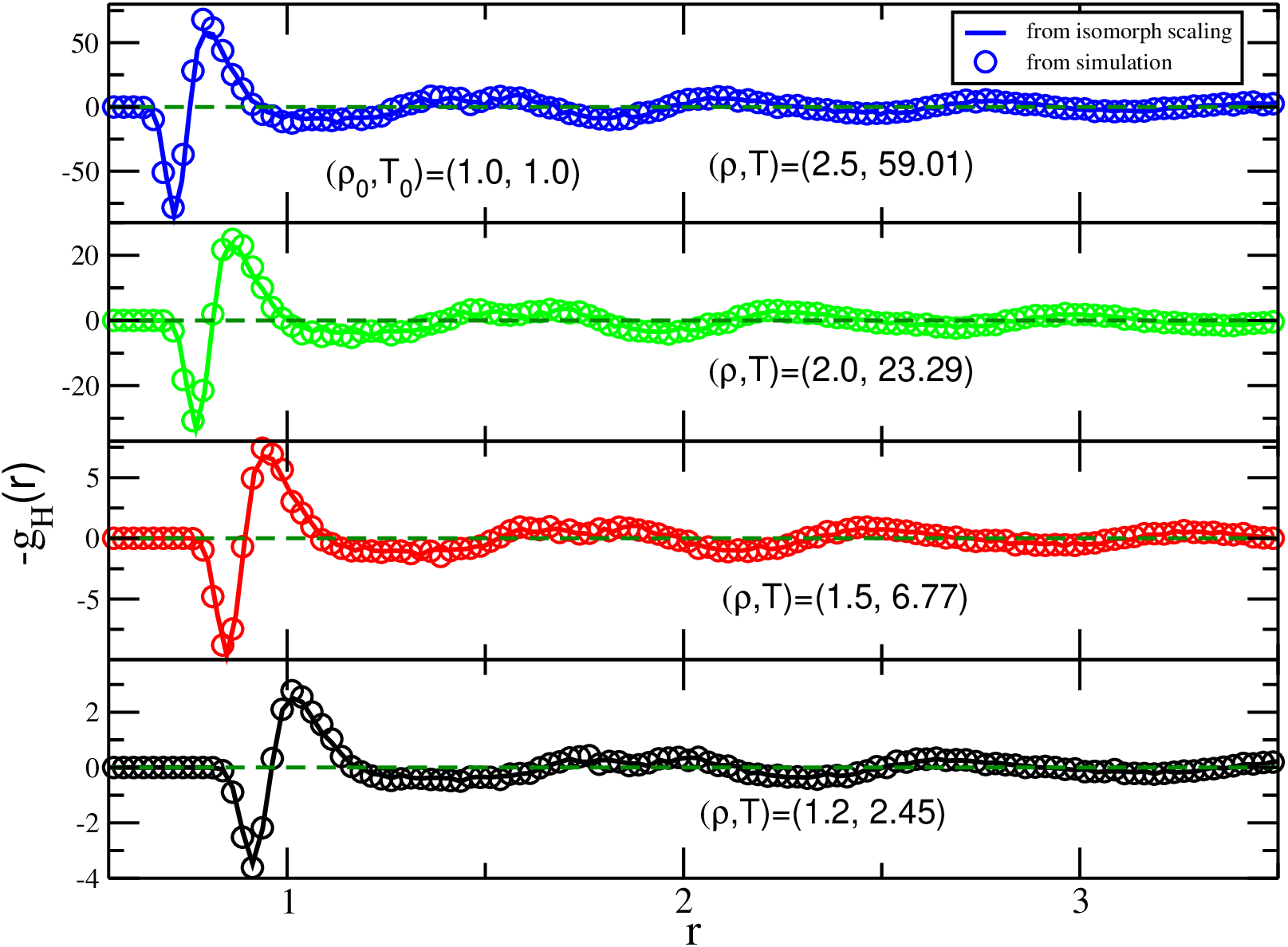}\vspace{1mm}
	\caption{A comparison of $-g_H(r)$ obtained from isomorph scaling and direct simulation at different state points on the isomorph.}
\label{gH-iso_app}
\end{figure}

Fig.~\ref{gH-iso_app} is similar to Fig.~\ref{gH-iso}, but for the isomorph in the supercooled regime.
The $-g_H(r)$ obtained from isomorph scaling and simulations are in great agreement, similar to along the isomorph in the normal liquid regime (Fig.~\ref{gH-iso}).

\begin{figure*}
	\includegraphics[width=0.40\textwidth,angle=0]{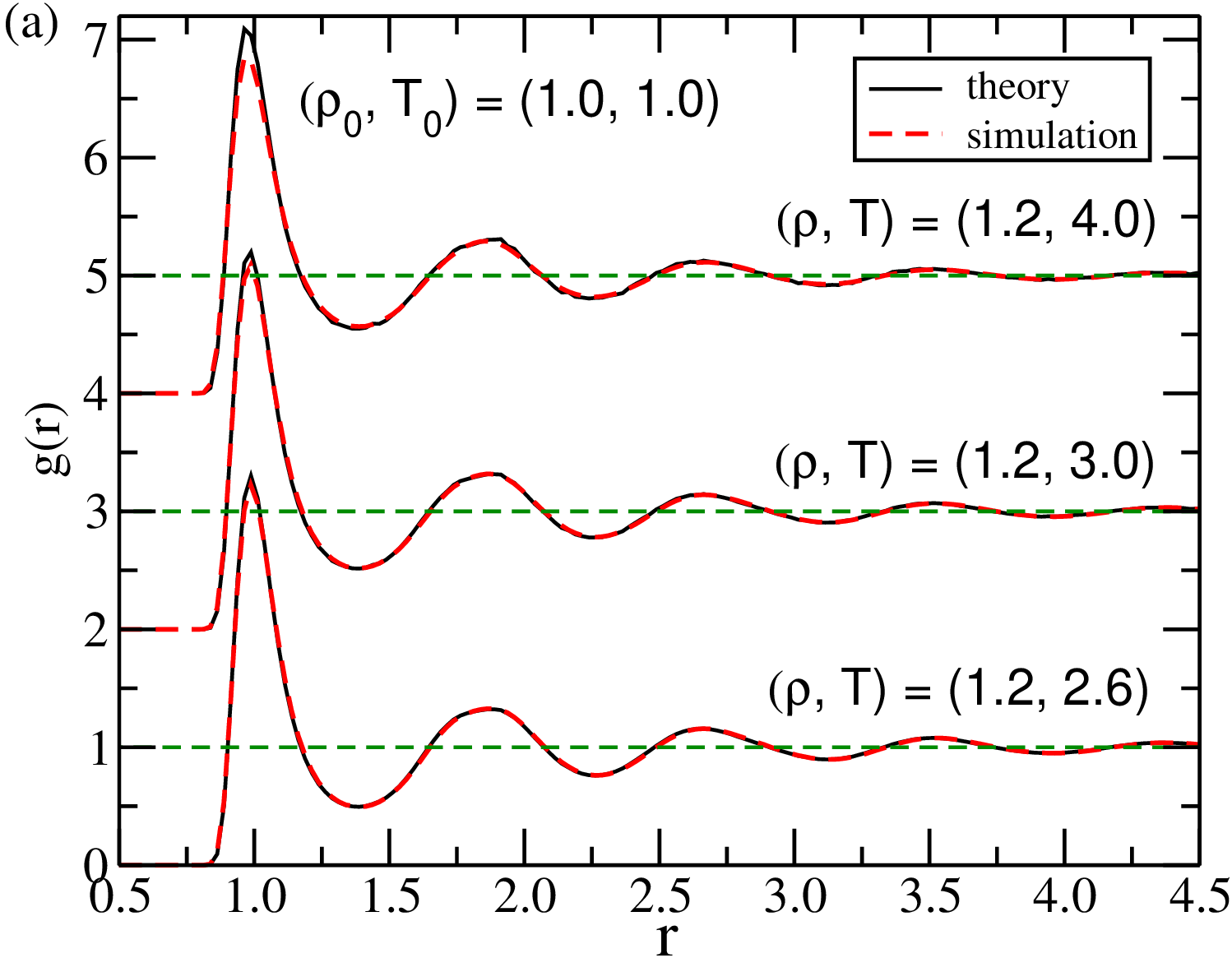}\hspace{9mm}\includegraphics[width=0.40\textwidth,angle=0]{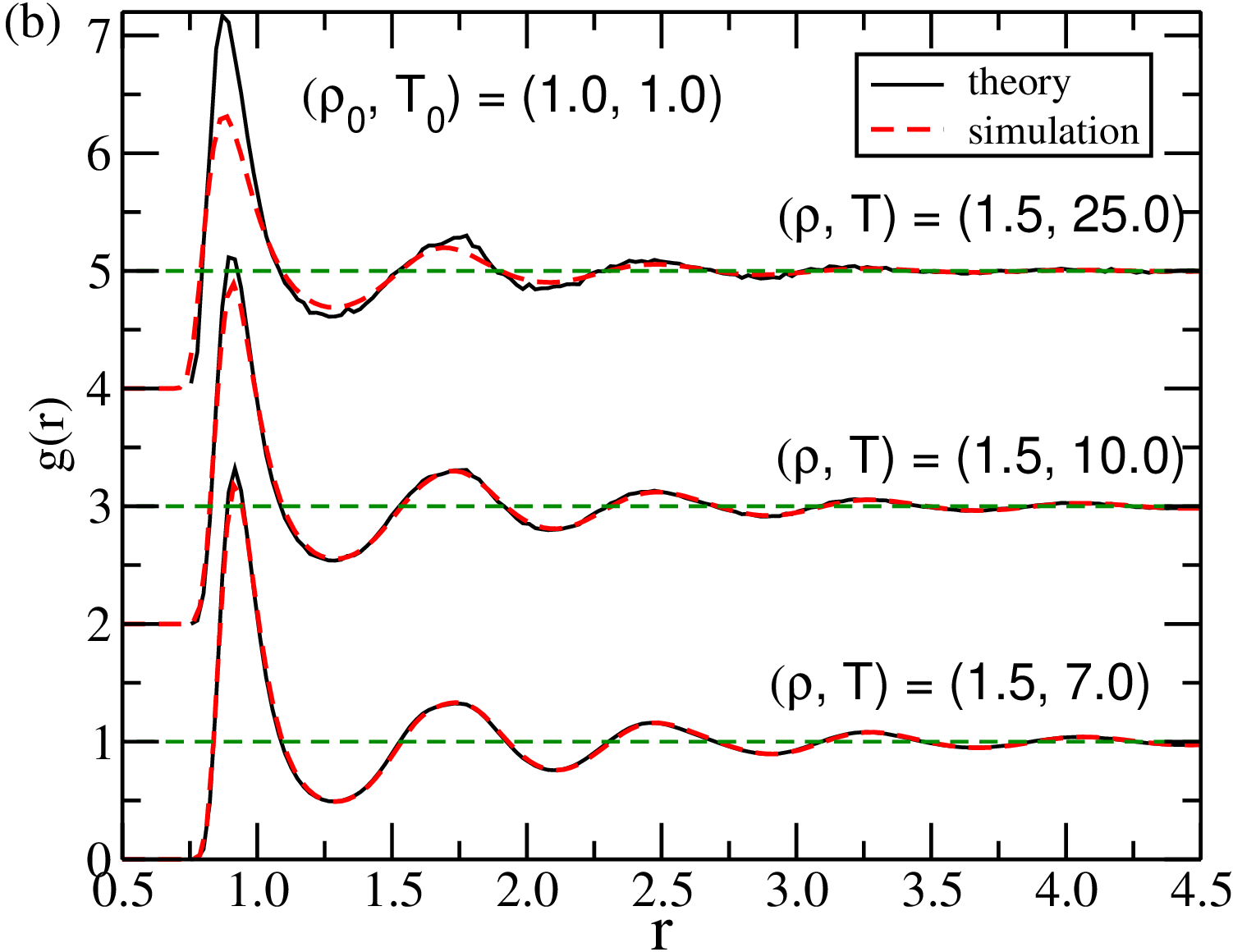}\vspace{9mm}
	\includegraphics[width=0.40\textwidth,angle=0]{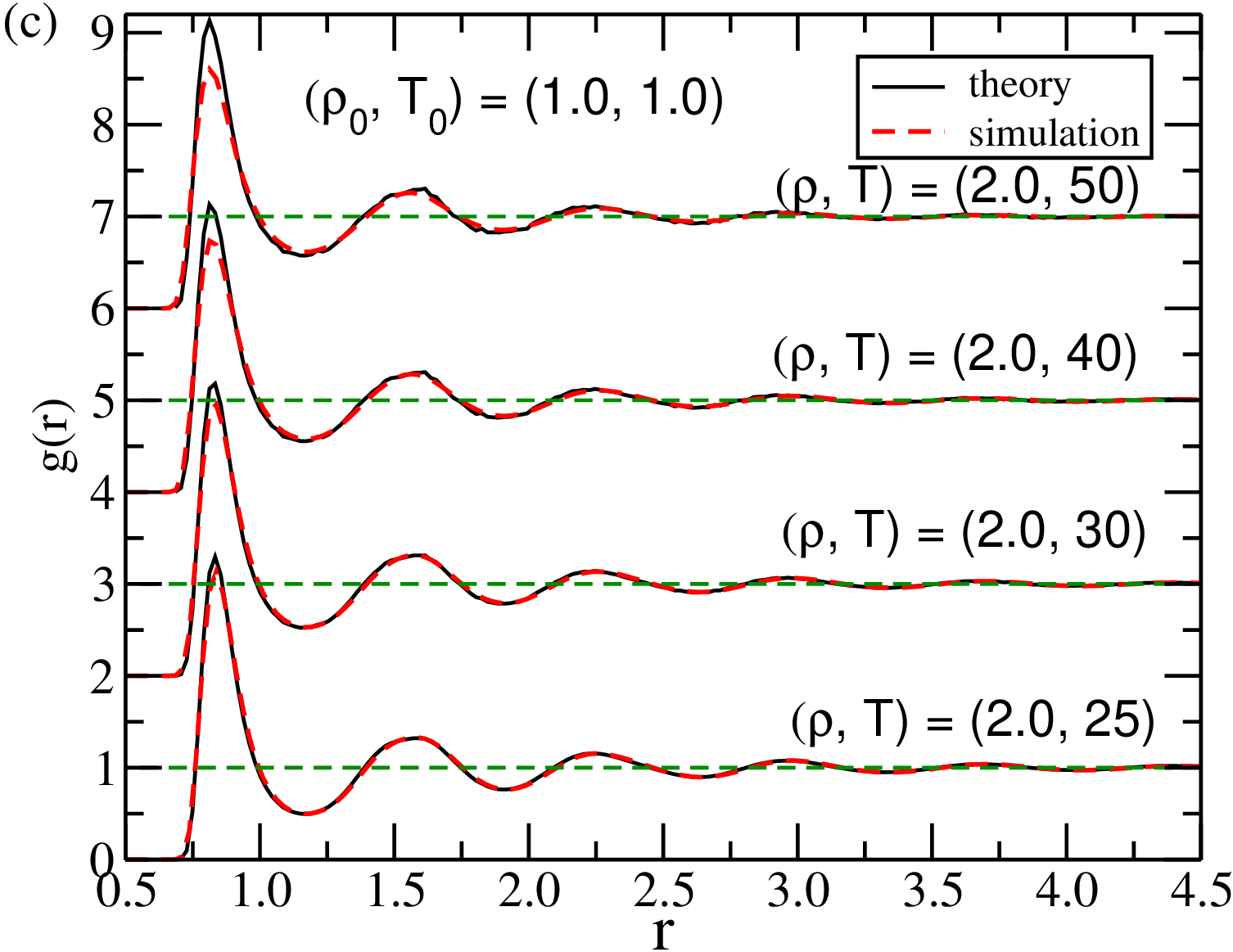}\hspace{9mm}\includegraphics[scale=0.25,angle=0]{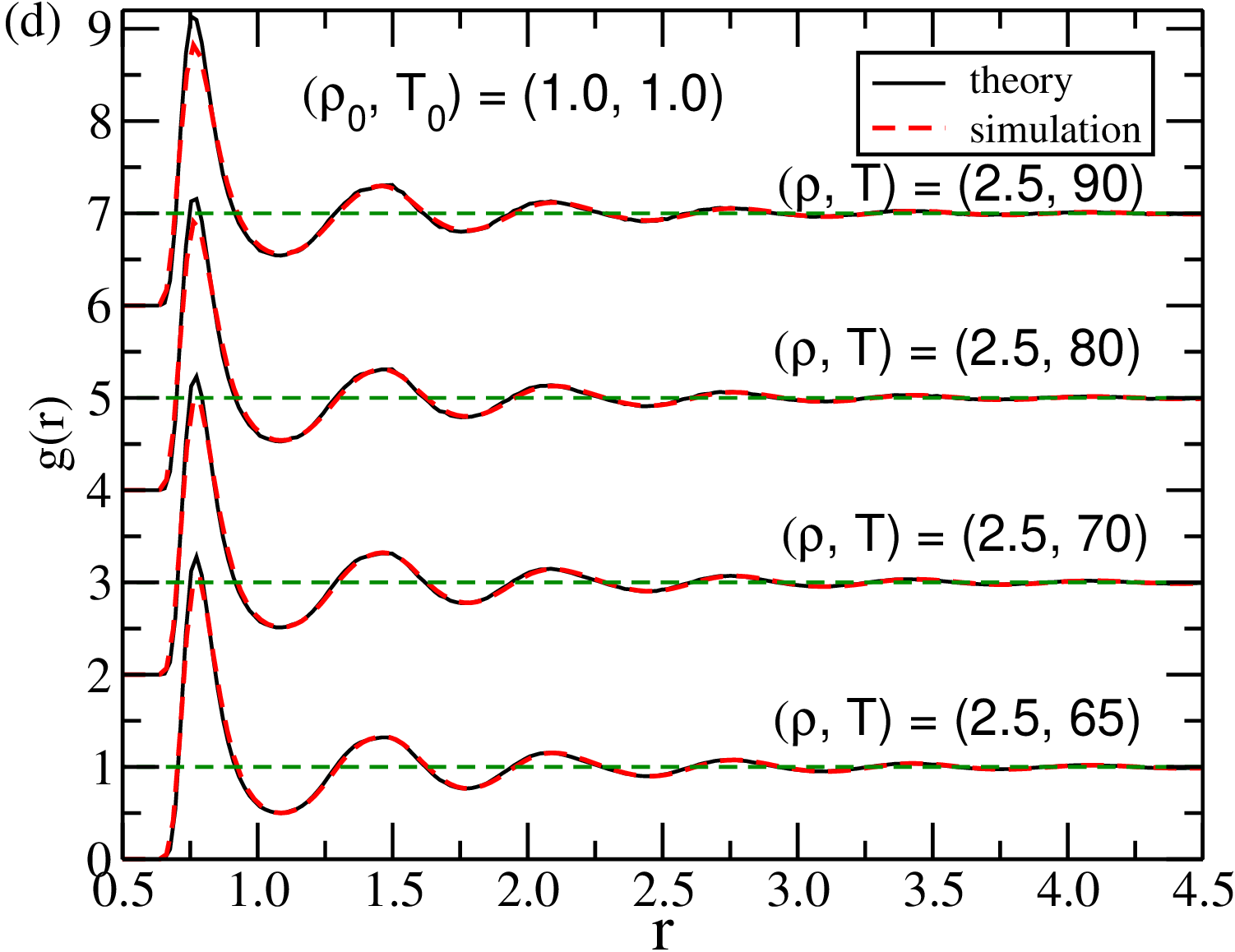}\vspace{0mm}
	\caption{A comparison of $g(r)$ obtained by employing Piskulich-Thompson+isomorph theory with that from simulation at different points of isochores (a) $\rho=1.20$ (b)$\rho=1.50$ (c)$\rho=2.00$, and (d)$\rho=2.50$. The reference temperature and density for Piskulich-Thompson+isomorph theory is ($\rho_0=1.0, T_0=1.0$).}
\label{gr_app}
\end{figure*}

Fig.~\ref{gr_app} is similar to  Fig.~\ref{iso-rho1.1-2.5} except that the reference state point $(\rho_0=1, T_0=1)$, where a single simulation is performed, is now in the supercooled 
regime. Fig.~\ref{gr_app} shows that the Piskulich-Thompson+isomorph theory is able to determine the $g(r)$ at an arbitrary state points below the van't Hoff demarcation line.

\end{document}